%% file: R1_JMVA_Submission.tex
\documentclass[authoryear,12pt]{article}

\usepackage{amsthm}
\usepackage{color}
\usepackage[latin1]{inputenc}
\usepackage{amsfonts}
\usepackage{amssymb,bm, bbm}
\usepackage{amsmath,enumerate,rotate}
\usepackage{amssymb,latexsym,mathrsfs}
\usepackage{algorithm}
\usepackage{epsfig,setspace}
\usepackage{graphicx}
\usepackage{caption}
\usepackage[english]{babel}
\usepackage{natbib}
\usepackage{color}
\usepackage{multirow}
\usepackage{booktabs}

\theoremstyle{theorem}
\newtheorem{theorem}{Theorem}
\theoremstyle{theorem}

\theoremstyle{definition}
\newtheorem{definition}{Definition}

\usepackage[top=1in, bottom=1in, left=1in, right=1in]{geometry}
\input{definitions.tex}

\usepackage{DejaVuSans}

\begin{document}

\title{Compound vectors of subordinators and their associated positive L\'evy copulas}

\author{
\hspace{-20pt}
Alan Riva Palacio\textsuperscript{a} \hspace{10pt}
Fabrizio Leisen\textsuperscript{b}  \hspace{10pt}\hspace{10pt} 
 \\\\
       {\centering {\small
       \textsuperscript{a}Universidad Nacional Aut\'onoma de M\'exico \hspace{5pt} \textsuperscript{b} University of Nottingham, U.K.}} \vspace{5pt} \\
     }

\date{}

\maketitle

\abstract{\noindent L\'evy copulas are an important tool which can be used to build dependent L\'evy processes.  In a classical setting, they have been used to model financial applications. In a Bayesian framework they have been employed to introduce dependent nonparametric priors which allow to model heterogeneous data. This paper focuses on introducing a new class of L\'evy copulas based on a class of subordinators recently appeared in the literature, called \textit{Compound Random Measures}. The well-known Clayton L\'evy  copula is a special case of this new class. Furthermore, we provide some novel results about the underlying vector of subordinators such as a series representation and relevant moments. The article concludes with an application to a Danish fire dataset.
}

\bigskip


%
%
%
%
%
%
%
%
%
%
%
%
%
%
%

\section{Introduction}\label{sec:intro}

\noindent Vectors of subordinators, namely a real valued non-decreasing stochastic process with independent increments, are an important class of processes which have been used for the modeling of data arising from multiple components. For example, \cite{yuen} solve the ruin problem for a bivariate Poisson process, \cite{semeraro} uses a vector of Gamma processes to construct a multivariate variance gamma model for financial applications and \cite{esma} perform parameter estimation for bivariate compound Poisson processes which they apply to an insurance dataset. \cite{esma2} focus on parameter estimation for a vector of stable processes, \cite{jiang} deal with a vector of gamma processes, and \cite{esma3} present a two-step estimation method for general multivariate L\'evy processes. In the context of Bayesian non-parametric statistics, vectors of subordinators have been used to construct dependent priors to model heterogeneous data; the celebrated Dirichlet Process, introduced in \cite{ferguson}, can be seen as a normalized Gamma subordinator. In the context of survival analysis, \cite{doksum} employs  1-dimensional subordinators to build the so-called \textit{neutral to the right} priors. More complex Bayesian nonparametric priors based on vectors of subordinators have been proposed such as, the vectors of dependent random measures in \cite{lijoi}, obtained through entrywise normalization of a vector of subordinators or the multivariate survival priors in \cite{epifani} and \cite{RPL2018a} which use vectors of subordinators to extend the neutral to the right priors into a \textit{partially exchangeable setting}. \cite{ishwaran} proposed a vector of generalized gamma processes and \cite{leisenlijoi} proposed a vector of Poisson-Dirichlet processes constructed using a vector of stable processes. More recently \cite{camer1} and \cite{camer2} proposed flexible dependent priors to model data heterogeneity. 
\\
\\
\noindent
The dependence structure for the entries of a vector of subordinators is particularly important for application purposes. In this context, the main approach to model the dependence is the one of \textit{L\'evy copulas} where in analogy with \textit{distributional copulas}, see \cite{nelsen}, the marginal behavior of the vector of subordinators can be decoupled from the dependence structure.
As highlighted in \cite{tankovrev}, L\'evy copulas have found important applications in statistical inference for vectors of \textit{L\'evy processes}, the study of multivariate regular variation and risk management applications. In Section 4 we will focus on parameter inference  as discussed in \cite{esma}, \cite{esma2} and \cite{esma3}. 
 
In a Bayesian non-parametric framework \cite{GL2017} introduced a class of vector of subordinators which relies on a one-dimensional subordinator and a $d$-variate probability distribution in $\left( \re^+ \right)^d$ to determine the corresponding  vector. Although their construction relies on the concept of \textit{completely random measures}, see \cite{kingmancrm}, in this work we use the equivalent setting of subordinators and present new results regarding these vectors which henceforward, we call \textit{compound vectors of subordinators}. In particular, we present a novel compound vector of subordinators which exhibits asymmetry in its related L\'evy copula. We provide a series representation for compound vectors of subordinators and exemplify its use for simulation purposes. We also provide  a new criteria for compound vector of subordinators to be well defined and give formulas for the associated fractional moments of order less than one, means, variances and covariances.  \cite{GL2017} showed the structure of the L\'evy copula associated to a compound vector of subordinator in a particular case, namely when what they call the score distribution in their construction has  independent and identically distributed marginal distributions; further exploration of the L\'evy copula structure was not performed. In the present work we explore the general L\'evy copula structure associated to a vector of compound subordinators. On the other hand, we give a tractable example of an asymmetric family of L\'evy copulas arising from a compound vector of subordinators. This new family is interesting as it contains the symmetric Clayton L\'evy copula as a particular case and preserves the behavior of a parameter modulating between indepence and complete dependence while allowing an extra parameter $\pmb{\alpha}$ to modulate asymmetry. In a similar fashion to \cite{esma}, we show the use of such new family for the modelling of a bivariate compound Poisson process and the related parameter inference. A simulation study for a vector of stable processes and a real data study pertaining insurance are performed. We conclude that broader classes of L\'evy copulas are needed and the compound vector of subordinators approach is a valid and tractable way to do so.  

\medskip

\noindent The outline of the paper is the following. Section 2 introduces vectors of dependent subordinators and L\'evy copulas. Section 3 and Section 4 are devoted to illustrate the main results of the paper. Section 5 includes application of the new model to simulated and real data sets. Section 6 concludes with a discussion. All the proofs of the results are in the Appendix.
 
\section{Preliminaries}
\noindent This section is devoted to introduce some preliminary notions about vectors of subordinators and L\'evy copulas. 
\begin{definition}
We say that $\pmb{Y}=(Y_1,\ldots ,Y_d)$, $d\in \mathbb{N},$ is a $d$-variate vector of positive jump L\'evy processes if for $t>0$, $\pmb{\lambda}\in (\re^+)^d$
\begin{align*}
\esp{e^{-\lambda_1 Y_1(t) - \ldots - \lambda_d Y_d (t) }}
&=e^{-\int_{ (\re^+)^d \times [0,t] }
(1-e^{- \lambda_1 s_1 - \ldots - \lambda_d s_d })\nu( \d s_1,\ldots , \d s_d , \d x)
}
\\
&=
e^{-\int_{ (\re^+)^d \times [0,t] }
(1-e^{- \langle \pmb{\lambda}\, , \,  \pmb{s} \rangle })\nu( \d \pmb{s} , \d x)
},
\end{align*}
with $\nu$ a measure in $\left( (\re^+)^{d+1} , \mathcal{B}\left( (\re^+)^{d+1} \right) \right)$ such that
\begin{align}\label{integcond}
\int_{(\re^+)^d \times \re^+ }
\min \left\{
1, \| \pmb{s} \|
\right\}
\nu(\d \pmb{s}, \d x ) < \infty .
\end{align}
We call $\nu$ the L\'evy intensity of $\pmb{Y}$.
\end{definition}
\noindent
In the following we refer to the stochastic process defined above as a vector of subordinators. 
We say that a L\'evy intensity is homogeneous if 
$$
\nu(\d \pmb{s} , \d x)=\rho(\d \pmb{s})\alpha (\d x).
$$
We define the Laplace exponent of an univariate subordinator, see \cite{sato} for details.
\begin{definition}\label{lapexpdef}
Let $\nu$ be a L\'evy intensity with $d=1$ and associated subordinator $Y$. We say that the Laplace exponent of $\nu$ is 
$$
\psi_t (\lambda)=\int_0^t \int_0^\infty (1-e^{-\lambda s})\nu(\d s, \d x)=-\log\left( \esp{e^{-\lambda Y(t) } } \right).
$$
\end{definition}
\noindent

\noindent
The tail integral of a vector of subordinators plays an important role in the results displayed in Section 3 and 4. It is defined as follows. 
\begin{definition}
Let $\pmb{Y}$ be a vector of subordinators with homogeneous L\'evy intensity $\rho$. Its associated \textit{tail integral} is defined as
\begin{equation}
U(\pmb{y})= \int_{ [y_1,\infty) \times  \ldots \times 
[ y_d , \infty ) } \rho (\d \pmb{s} ).
\end{equation}
\noindent
The marginal tail integrals associated to $U(\pmb{y})$ are given by
\begin{equation*}
U_i(y)=U(y_1^{(i)}, \ldots , y_{i-1}^{(i)}, y, y_{i+1}^{(i)},\ldots, y_d^{(i)}),
\end{equation*}
where $y_1^{(i)}= \cdots = y_{i-1}^{(i)}= y_{i+1}^{(i)},=\ldots = y_d^{(i)} = 0$ for $i \in \{ 1, \ldots , d \}$.
\end{definition}
\noindent
Given a vector of subordinators, $\pmb{Y}=(Y_1, \ldots , Y_d)$, there exist collections of random elements $\{W_{1,i}\}_{i=1}^\infty, \ldots ,\{W_{d,i}\}_{i=1}^\infty$ and $\{V_i\}_{i=1}^\infty$ such that
\begin{align}\label{serrepr}
\left( Y_1 (t) , \ldots , Y_d(t) \right) & \stackrel{\text{a.s.}}{=}
\left( 
\sum_{i=1}^\infty W_{1,i} \indic_{ \left\{ V_i \leq t \right\}}
, \ldots , 
\sum_{i=1}^\infty W_{d,i} \indic_{ \left\{ V_i \leq t \right\}}
\right).
\end{align}
For a full review of vectors of subordinators see \cite{tankbook}. The construction in \cite{GL2017} for vectors of completely random measures can be set in the context of vectors of subordinators as follows.
\begin{definition}
Let $h$ be a $d-$variate probability density function and $\nu^\star$ a univariate L\'evy intensity. We say that a vector of subordinators $\pmb{Y}$ is a $d-$variate compound vector of subordinators with score distribution $h$ and directing L\'evy measure $\nu^\star$ if it has a $d-$variate L\'evy intensity given by
\begin{equation*}
\nu(\d \pmb{s}, \d x )= \int z^{-d}h(s_1/z, \ldots , s_d/z) \nu^\star (\mathrm{d}z, \d x ) \mathrm{d}\pmb{s}.
\end{equation*}
\end{definition}
\noindent If the directing L\'evy measure is homogeneous, $\nu^\star (\d \pmb{s},\d x)=\rho^\star (\d \pmb{s})\alpha (\d x)$, then $\nu(\d \pmb{s}, \d x )=\rho (\d \pmb{s} )\alpha (\d x) $ with
$$
\rho (\d \pmb{s} ) =
\int z^{-d}h(s_1/z, \ldots , s_d/z) \rho^\star (\mathrm{d}z ) \mathrm{d}\pmb{s}.
$$ 
In the next result we present what will be the running working example of this work. In particular, we restrict ourselves to the $2$-dimensional vector of subordinators setting for illustration purposes.
\begin{theorem}\label{workexpteo}
Let $\sigma \in (0,1)$, and $\alpha_1,\beta_1,\alpha_2,\beta_2>0$. If the score distribution is given by
\begin{equation*}
h(y_1,y_2)=\frac{ \beta_1^{\alpha_1}\beta_2^{\alpha_2} }{\Gamma(\alpha_1)\Gamma(\alpha_2)}
y_1^{\alpha_1-1} e^{-\beta_1y_1} y_2^{\alpha_2-1}e^{-\beta_2y_2},
\end{equation*}
i.e. the score distribution is given by independent $\text{Gamma}(\alpha_1,\beta_1)$, $\text{Gamma}(\alpha_2,\beta_2)$ distributions, and the directing L\'evy measure has intensity
\begin{equation*} 
\rho^\star(z)=\sigma K z^{-\sigma-1},
\end{equation*}
i.e. a $\sigma$-stable intensity with proportionality parameter $K=\sigma/\Gamma(1-\sigma)$. Then the corresponding compound vector of subordinators has a bivariate L\'evy intensity given by
\begin{align}\label{workexintens}
\rho_{\sigma,K,\pmb{\alpha},\pmb{\beta}}
(\d s_1,\d s_2)
=
\frac{\sigma K \beta_1^{\alpha_1}\beta_2^{\alpha_2}
\Gamma(\alpha_1 +\alpha_2 +\sigma)s_1^{\alpha_1 -1}s_2^{\alpha_2 - 1}
}{\Gamma (\alpha_1)\Gamma (\alpha_2)(\beta_1 s_1 +\beta_2 s_2)^{\alpha_1+\alpha_2+\sigma}}\d s_1 \d s_2,
\end{align}
and has $\sigma$-stable marginals with proportionality parameters
\begin{align*}
K_i=\frac{ K \beta_i^{-\sigma} 
\Gamma(\alpha_i+\sigma)
}{\Gamma (\alpha_i)},
\end{align*}
corresponding to each dimension $i\in\left\{1,2\right\}$.
\end{theorem}
\noindent
The above result is a generalization of Corollary 1
in \cite{GL2017} where $\alpha_1=\alpha_2$ and $\beta_1=\beta_2=1$ was considered. Furthermore, the case $\alpha_1=\alpha_2=\beta_1=\beta_2=1$ was considered by \cite{epifani} and \cite{leisenlijoi} who link it to Clayton L\'evy copulas, to be discussed next. This simple example exhibits the possibility to tractably link compound vectors of subordinators to L\'evy copulas. In \cite{GL2017} only symmetric multivariate L\'evy intensities where considered, however departure from such symmetry is of importance for modelling purposes as will be showed in Section 5. In Section 4 we will present the asymmetric L\'evy copula associated to the particular compound vector of subordinators above. 
\\
\\
A popular approach for modelling the dependence structure of vectors of subordinators is given by L\'evy copulas.
\begin{definition}
A $d-$variate positive L\'evy copula is a function $\mathcal{C}(s_1, \ldots , s_d):[0,\infty]^d\rightarrow [0,\infty]$ which satisfies
\begin{enumerate}
\item $\mathcal{C}(s_1, \ldots , s_d)<\infty$ for $(s_1, \ldots , s_d)\neq (\infty, \ldots , \infty)$.
\item $\mathcal{C}$ is $d-$increasing.
\item $\mathcal{C}(s_1, \ldots , s_d)=0$ if $s_k = 0$ for any $k \in \{1, \ldots , d\}$
\item $\mathcal{C}(y_1^{(k)},\dots,y_{k-1}^{(k)},s,y_{k+1}^{(k)},\dots,y_d^{(k)})=s$ for $k\in\{1,\dots ,d\}$, $s\in \re^+$, where
$y_1^{(k)}=\dots=y_{k-1}^{(k)}=y_{k+1}^{(k)}=\dots =y_d^{(k)}=\infty$. 
\end{enumerate}
\end{definition}
\noindent 
Such L\'evy copulas can be linked to a vector of subordinators via the following theorem.
\begin{theorem}[\textbf{\cite{tankbook}}]\label{sklarlevycop}
\textbf{(Sklar's Theorem for tail integrals and  L\'evy copulas)}
Let $U$ be a $d$-variate tail integral
with margins $U_1, \ldots , U_d$ then there exists a L\'evy copula $\mathcal{C}$ such that
\begin{equation*}
U(\pmb{y})=\mathcal{C}(U_1(y_1),\ldots , U_d(y_d)).
\end{equation*}
If $\{U_i\}_{i=1}^d$ are continuous $\mathcal{C}$ is unique, otherwise it is unique in $\text{Ran}(U_1)\times \ldots \times \text{Ran}(U_d)$.
\end{theorem}
\noindent
For a proof see Theorem 5.3 in \cite{tankbook}. If the L\'evy copula is smooth enough then from Theorem \ref{sklarlevycop} and the definition of the tail integral we have that the underlying multivariate L\'evy intensity can be expressed as
\begin{align}\label{copmultivintensexpr}
\rho (\pmb{s})=\restr{\frac{\partial^d}{\partial u_1\cdots \partial u_d}\mathcal{C}(\pmb{u})}{u_1=U_1(s_1),\cdots , u_d=U_d(s_d)}\rho_1(s_1)\cdots \rho_d(s_d),
\end{align}
where $\rho_i$, $i\in \{ 1, \ldots , d\}$, are the corresponding marginal L\'evy intensities associated to the tail integrals $U_1, \ldots , U_d$. Furthermore if $\mathcal{C}$ is a two dimensional L\'evy copula and $\{ \left( W_{1,i} , W_{2,i} \right) \}_{i=1}^\infty$ are the random weights of a series representation for the associated vector of subordinators, equation \eqref{serrepr}, then the law of $S_{1,i} = U_1 \left( W_{1,i} \right)$ conditioned on $S_{2,i} = U_2 \left( W_{2,i} \right)=s_2 \in \re^+\setminus \{ 0 \}$ is given by the distribution function
\begin{align}\label{copula_cond1}
\hat{F}_{S_1|S_2=s_2}(s_1) & = 
\frac{\partial}{\partial s_2} \mathcal{C}(s_1, s_2),
\end{align}
and the law of $S_{2,i}=U_2 \left( W_{2,i} \right)$ conditioned on $S_{1,i}=U_1 \left( W_{1,i} \right) =s_1 \in \re^+ \setminus \{ 0\}$ is given by the distribution function
\begin{align}\label{copula_cond2}
\hat{F}_{S_2|S_1=s_1}(s_2) & = 
\frac{\partial}{\partial s_1} \mathcal{C}(s_1, s_2);
\end{align}
see Theorem 6.3 in \cite{tankbook} for a proof. Some examples of $d$-variate positive L\'evy copulas are the following:
\begin{ex}\label{indeplevycop}
\textbf{Independence L\'evy copula.} 
$$\mathcal{C}_{\perp}(s_1, \ldots ,s_d)=
\sum_{i=1}^d s_i \prod_{j\neq i} \indic_{\{ s_j = \infty \}}
.$$ In this case the subordinators $Y_1, \ldots ,Y_d$ are pairwise independent.
\end{ex}
\begin{ex}\label{compdeplevycop}
\textbf{Complete dependence L\'evy copula.} 
$$\mathcal{C}_{||}(s_1,, \ldots ,s_d) = \min\{s_1,\ldots ,s_d\}.$$ 
In this case the subordinators $Y_1, \ldots , Y_d $ are completely dependent in the sense that the vector of jump weights for the associated series representation, \eqref{serrepr},  $ \left\{ \left( W_{1,i}, \ldots , W_{d,i} \right) \right\}_{i=1}^\infty$, are in a set $S$ such that whenever $\pmb{v},\pmb{u}\in S$ then either $v_j<u_j$ or $u_j<v_j$ for all $j\in\{1,\dots ,d\}$.
\end{ex}
\noindent The following L\'evy copula example is of interest in the literature as it has as limiting cases the independence and complete dependence examples above.
\begin{ex}\label{claytlevycop}
\textbf{ Clayton L\'evy  copula.}
$$\mathcal{C}_\theta(s_1, \ldots ,s_d)=\left(
s_1^{-\theta}+\ldots + s_d^{-\theta}\right)^{-1/\theta};\quad \theta >0.$$
The parameter $\theta$ in the Clayton  L\'evy  copula allows us to modulate between the independence and complete dependence cases as
$$
\lim_{\theta \to 0 } \mathcal{C}_\theta
(s_1, \ldots ,s_d) =  \mathcal{C}_{\perp}(s_1, \ldots ,s_d)
$$
and
$$
\lim_{\theta \to \infty } \mathcal{C}_\theta
(s_1, \ldots ,s_d) =  \mathcal{C}_{||}(s_1,, \ldots ,s_d).
$$
\end{ex}
\noindent
Such example is the L\'evy copula analogue of the distributional Clayton copula which also modulates between independence and complete dependence cases for multivariate probability distributions, see \cite{nelsen}. We observe that the Clayton L\'evy copula is symmmetric which for real data applications can be too strong a constraint. For a full review of L\'evy copulas see \cite{tankbook}.
\section{Results for compound vectors of subordinators}
\noindent

\noindent This section provides general results for compound subordinators. In particular, we provide a series representation, conditions for the vector to be well posed and expressions for the mean, variance and correlation of the process. In the first result we provide a representation with the structure displayed in equation \eqref{serrepr}. 

\begin{theorem}\label{serreprteo}
Let $\pmb{Y}=(Y_1, \ldots ,  Y_d)$ be a compound subordinator given by a score distribution $h$ and directing L\'evy measure $\nu^\star$ with associated univariate subordinator $Y^\star$. Then for $\bm{t}\in \left( \re^+\right)^d$
\begin{align*}
\pmb{Y}(\pmb{t}) & = 
\left( Y_1(t_1), \ldots , Y_d(t_d) \right)
 \stackrel{\text{a.s.}}{=} \left(
\sum_{i=1}^\infty M_{1,i}W_i \indic_{ \left\{ V_i \leq t_1 \right\} }
, \ldots,  \sum_{i=1}^\infty M_{d,i} W_i \indic_{ \left\{ V_i \leq t_d  \right\}}
\right),
\end{align*}
where
\begin{equation*}
Y^\star (t) \stackrel{\text{a.s.}}{=} \sum_{i=1}^\infty W_i \indic_{ \left\{ V_i \leq t \right\} }
\end{equation*}
with $t\in \re^+$, and $$(M_{1,i},\ldots , M_{d,i})\stackrel{\text{i.i.d.}}{\sim}h.
$$
\end{theorem}
\begin{figure}
\begin{center}
\includegraphics[width=0.7\linewidth]{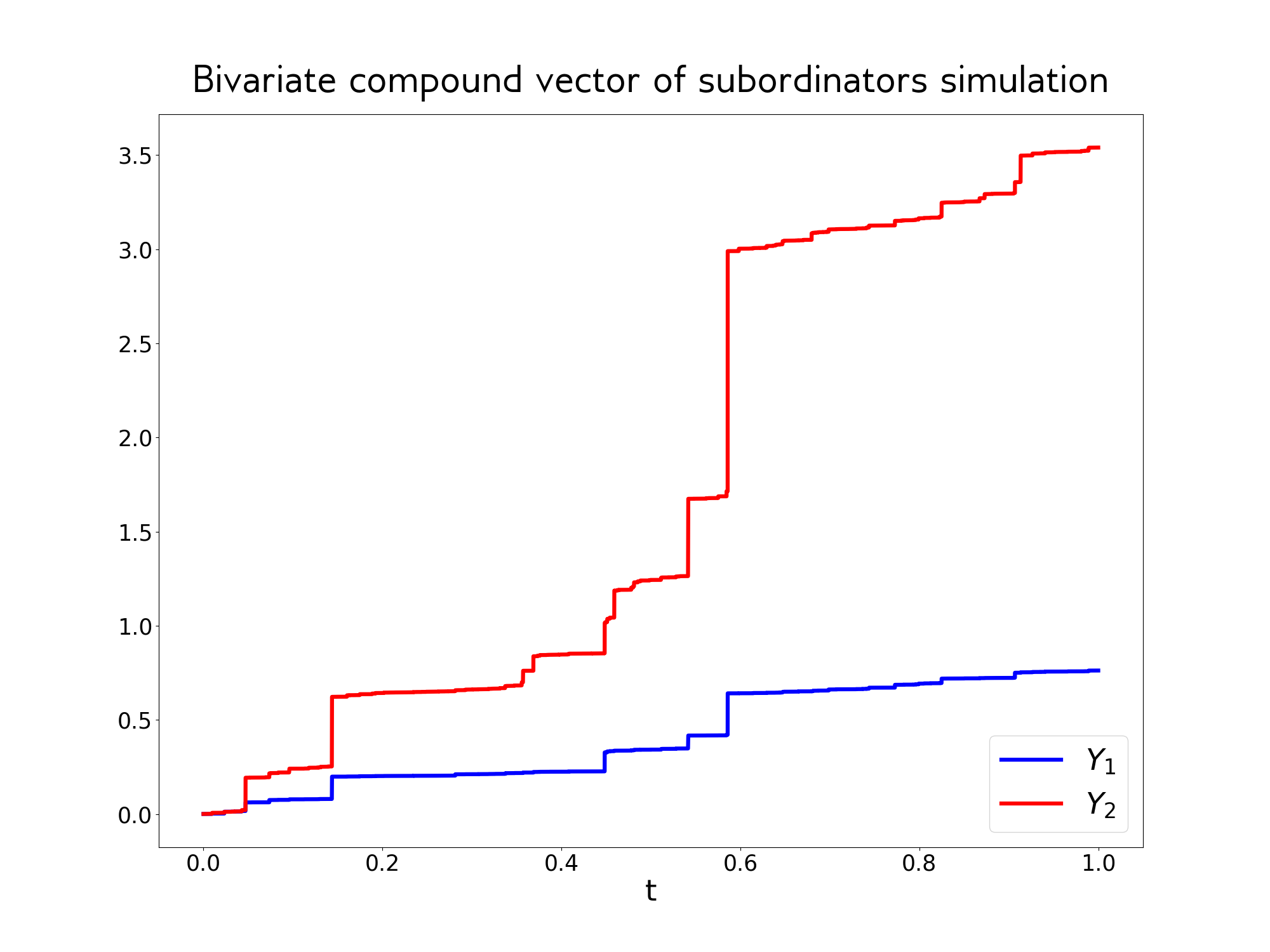}
\caption{Simulation in $[0,1]$ of working example compound vector of subrodinators $\pmb{Y}=(Y_1,Y_2)$ as in Theorem \ref{workexpteo} with $\alpha_1=1$, $\beta_1=2$, $\alpha_2=10$, $\beta_2=5$, $\sigma=0.5$ and $K=1$; obtained by the series representation \eqref{compserrepr} with weights associated to the directing L\'evy measure restricted to be greater than $\tau = 10^{-6}$.}\label{biv_sim}
\end{center}
\end{figure}
\noindent The above result is useful for computational purposes and provides a deeper understanding of the discrete structure of the process. In particular, if a series representation of the subordinator associated to the directing L\'evy measure is available then a simulation algorithm for the compound vector of subordinators can be constructed. Let $Y^\star$ and $U^\star$ be, respectively, the subordinator and tail integral associated to a directing L\'evy measure $\nu^\star$ which is absolutely continuous with respect to Lebesgue measure, and $(U^\star)^{-1}$ be the inverse of the tail integral. A popular series representation is given by the Ferguson-Klass algorithm as it follows.
\begin{theorem}\label{serreprteo}
Let $\left\{ U_i \right \}_{i=1}^\infty$ be i.i.d uniform random variables in $[0,1]$, $\left\{ T_i \right \}_{i=1}^\infty$ be i.i.d. standard exponential random variables and $\Gamma_k=\sum_{i=1}^k T_i$. Then 
\begin{equation*}
Y^\star (t) = \sum_{i=1}^\infty 
(U^\star)^{-1}(\Gamma_i) \indic_{ \left\{ U_i \leq t \right\} },
\end{equation*}
for $t\in[0,1]$.
\end{theorem}
\noindent
If the above series representation is truncated at an index $k$ such that $(U^\star)^{-1}(\Gamma_k)<\tau$ then the missing jump weights on the series are a.s. less than $\tau$. Letting $h$ be a score distribution and $\left\{ (M_{1,i},\ldots , M_{d,i})\right\}_{i=1}^k\stackrel{\text{i.i.d.}}{\sim}h$ then each entry of the compound vector of subordinators associated to $\nu^\star$ and $h$ has a truncated series approximation
\begin{equation}\label{compserrepr}
Y_j(t)\approx \sum_{i=1}^k M_{j,i}(U^\star)^{-1}(\Gamma_i)
\indic_{ \left\{ U_i \leq t \right\} },
\end{equation}
for $t\in [0,1]$ and $j\in \left\{1 ,\ldots , d\right\}$. Where the missing jump weights for entry $j$ are randomly bounded to be less than $\min\left\{ M_{j,1},\ldots M_{j,k} \right\}\tau$.
We refer to \cite{rosinski} for a full review of series representations for L\'evy processes and to \cite{tankbook} for a review of simulation algorithms for L\'evy processes. The inverse of the tail integral for a $\sigma$-stable L\'evy measure is readily available, see the proof of Theorem \ref{workexpteo} for details, so we can apply the Ferguson-Klass algorithm. In Figure \ref{biv_sim} we simulate a trivariate vector of subordinators given by our working example in Theorem \ref{workexpteo}. The next result provides practical condition to check if a compound vector of subordinators is well posed. 
\begin{theorem}\label{integcondteo}
Let $\nu^\star$ be a L\'evy measure and $h$ a $d$-variate score distribution such that if $(W_1, \ldots, W_d)\sim h$ then $\esp{W_i}<\infty$ $\forall i\in \{1, \ldots, d\}$. Then the compound vector of subordinators with directing L\'evy measure $\nu^\star$ and score distribution $h$ has a L\'evy intensity which satisfies condition \eqref{integcond}.
\end{theorem}
\noindent The above theorem improves the result presented in \cite{RPL2018b} by providing straightforward conditions to test if a vector of compound subordinators is well defined.
For instance, our working example in Theorem \ref{workexpteo}, with score distribution given by independent marginal Gamma distributions, can be readily seen to be well posed as Gamma random variables always have finite mean.
\\
This section concludes with a result which can be useful for modelling purposes and to understand the behavior of the process. Let $\psi_t$ be a Laplace exponent, we denote with $\psi_t'(0)=\restr{ \frac{\d}{\d \lambda}\psi_t(\lambda) }{\lambda = 0}$ the first derivative evaluated in $0$. The moments of a compound vector of subordinators are given in the next result. 

\begin{figure}
\begin{center}
\includegraphics[width=0.8\linewidth]{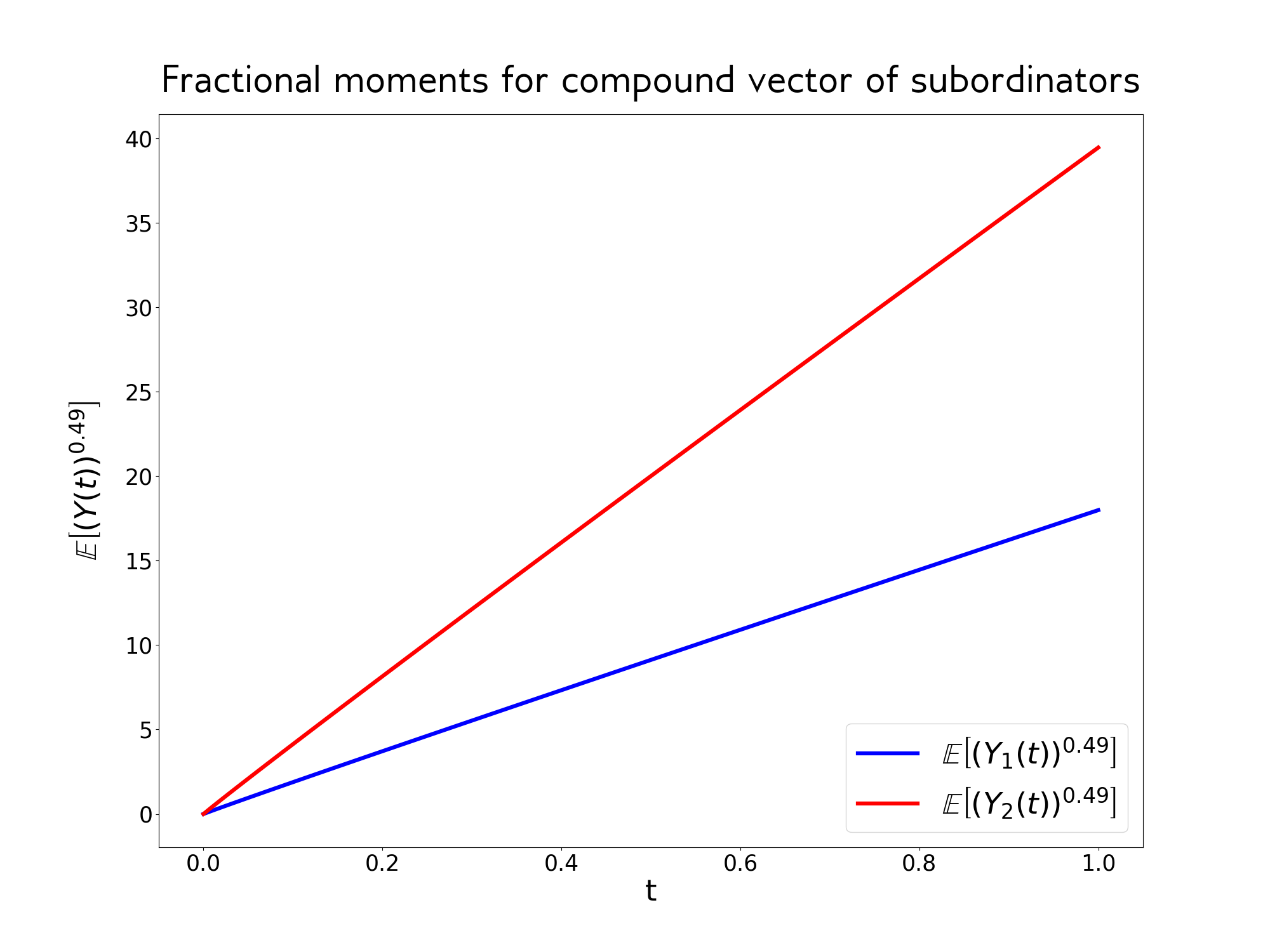}
\caption{Fractional moments of order $p=0.49$ for working example compound vector of subrodinators $\pmb{Y}=(Y_1,Y_2)$ as in Theorem \ref{workexpteo} with $\alpha_1=1$, $\beta_1=2$, $\alpha_2=10$, $\beta_2=5$, $\sigma=0.5$ and $K=1$ with $t\in[0,1]$.}\label{fract_mom_fig}
\end{center}
\end{figure}

\begin{theorem}\label{compoundmomentsteo}
Let $\pmb{Y}$ be a vector of compound subordinators with score distribution $h$ and directing L\'evy measure $\nu^\star$ with Laplace exponent $\psi_t^{\star}$ such that $\left( \psi_t^\star\right)'(0)$ and $\left( \psi_t^\star \right)''(0)$ exist $\forall t>0$; then
\begin{align*}
\esp{Y_i(t)}&=( \psi_t^\star )'(0)\esp{W_i},
\\
\esp{\left(Y_i(t)\right)^p}&=\frac{p}{\Gamma(1-p)}
\int_0^\infty 
\frac{1-e^{-\esp{\psi^\star_t(u W_i)}}}{u^{p+1}} \d u
\text{ for }p\in(0,1),
\\
\var{Y_i(t)} &= -( \psi_t^\star )''(0)\esp{W_i^2},
\\ \cov{Y_i(t),Y_j(t)} &= -( \psi_t^{\star} )''(0)\esp{W_i W_j},
\\ \corr{Y_i(t),Y_j(t)} &= \frac{
\esp{W_i W_j}
}{
\sqrt{\esp{W_i^2}\esp{W_j^2}}
},
\end{align*}
where $(W_1, \ldots, W_d)\sim h$ and $i,j\in \{1, \ldots , d\}$, $i\neq j$.
\end{theorem}
\noindent
The fractional moment formula of order $p\in(0,1)$ in the previous theorem is useful when dealing with $\sigma$-stable processes which do not have finite moments for $p\geq \sigma$. In our working example, the directing L\'evy measure is $\sigma$-stable which can be seen to have Laplace exponent $\psi^\star_t(\lambda)=K\lambda^\sigma$ and $W_i\sim \text{Gamma}(\alpha_i,\beta_i)$, which satisfies $\esp{W_i^\sigma}=\frac{\Gamma(\alpha_i +\sigma)}{\Gamma(\alpha_i)\beta_i^\sigma}$ with $i\in\left\{ 1,2\right\}$. So for $p<\sigma$ using the previous theorem and integrating by parts we have that
\begin{align*}
\esp{\left(Y_i(t)\right)^p}&=\frac{p}{\Gamma(1-p)}
\int_0^\infty 
\frac{1-e^{-\esp{ t K(u W_i)^\sigma }}}{u^{p+1}} \d u
=
\frac{K\esp{W_i^\sigma}}{\Gamma(1-p)}
\int_0^\infty 
\sigma e^{-tK\esp{ W_i^\sigma }u^\sigma}u^{\sigma-p-1} \d u
\\
&=
\frac{K\esp{W_i^\sigma}}{\Gamma(1-p)}
\int_0^\infty 
\sigma e^{-tK\esp{ W_i^\sigma }u}u^{1-\frac{p}{\sigma}-1} \d u  = \frac{ \left( tK\esp{W_i^\sigma} \right)^{\frac{p}{\sigma} } \Gamma(1-\frac{p}{\sigma}) }{\Gamma(1-p)},
\end{align*}
which agrees with Theorem 1 and the fractional moment formula for $\sigma$-stable processes, see \cite{sato} p. 162 or set $W_i\stackrel{\text{a.s.}}{=}1$ in the previous calculation. In Figure \ref{fract_mom_fig} we plot fractional moments for the working example of Theorem \ref{workexpteo}.

\section{Positive L\'evy copulas from compound vectors of subordinators}
\noindent 
In this section we provide a new family of L\'evy copulas, which has the  Clayton L\'evy copula in Example \ref{claytlevycop} as a particular case, and give a general formula for the L\'evy copula associated to a compound vector of subordinators. 

\subsection{ $(\alpha_1,\alpha_2)$-Clayton L\'evy copulas }
We present the family of $(\alpha_1,\alpha_2)$-Clayton L\'evy copulas. This new family allows for  asymmetry of the L\'evy copulas and has two extra parameters with respect to the Clayton L\'evy  copula, which offer more flexibility in modelling. We derive the new family by considering the L\'evy Copula associated to the compound vector of subordinators in Theorem \ref{workexpteo}.
Let the regularized incomplete beta function be given by
\begin{equation*}
I(x,\alpha,\beta)= \frac{\Gamma(a+b)}{\Gamma(a)\Gamma(b)} \int_0^x z^{\alpha-1}(1-z)^{\beta-1}\d z .
\end{equation*} 
The next theorem provides the L\'evy copula associated to the L\'evy intensity  in equation \eqref{workexintens}. 
\begin{theorem}\label{claylevycopteo}
Let $\sigma \in (0,1)$, $K>0$ and $\pmb{\alpha}=(\alpha_1, \alpha_2)$, $\pmb{\beta}=(\beta_1, \beta_2)$ have positive non-zero entries. Then
\begin{enumerate}[a)]
\item The L\'evy copula associated to $\rho_{\sigma,K,\pmb{\alpha},\pmb{\beta}}$ in Theorem \ref{workexpteo} is given by
\begin{align*}
\mathcal{C}_{\sigma,\pmb{\alpha}}(s_1, s_2) &=
s_1\,
I\left(
\frac{ \left( \frac{\Gamma(\alpha_1+\sigma)}{\Gamma(\alpha_1)s_1} \right)^{\frac{1}{\sigma}} }{ 
\left( \frac{\Gamma(\alpha_1+\sigma)}{\Gamma(\alpha_1)s_1} \right)^{\frac{1}{\sigma}} +
\left( \frac{\Gamma(\alpha_2+\sigma)}{\Gamma(\alpha_2)s_2} \right)^{\frac{1}{\sigma}}
},\alpha_1+\sigma,\alpha_2 \right)
\\
&\hphantom{=}+
s_2\,
I\left(
\frac{ \left( \frac{\Gamma(\alpha_2+\sigma)}{\Gamma(\alpha_2)s_2} \right)^{\frac{1}{\sigma}} }{ 
\left( \frac{\Gamma(\alpha_1+\sigma)}{\Gamma(\alpha_1)s_1} \right)^{\frac{1}{\sigma}} +
\left( \frac{\Gamma(\alpha_2+\sigma)}{\Gamma(\alpha_2)s_2} \right)^{\frac{1}{\sigma}}
},\alpha_2 +\sigma, \alpha_1 \right).
\end{align*}
\item Furthermore the above L\'evy copula $C_{\sigma,\pmb{\alpha}}$ can be extended for $\sigma \in (0,\infty)$.  
\end{enumerate}
\end{theorem}
\noindent We denote the L\'evy copulas in the above  theorem as $(\alpha_1,\alpha_2)$-Clayton or $\pmb{\alpha}$ Clayton L\'evy copulas. We highlight that although the L\'evy copula induced by $\rho_{\sigma,K,\pmb{\alpha},\pmb{\beta}}$ is only defined for $\sigma \in (0,1)$, b) in the above theorem tells us that such copula is still a copula when considering $\sigma \geq 1$. Although this may seem surprising, the following observation tells us that for $\pmb{\alpha}=(1,1)$ the $\pmb{\alpha}$-Clayton L\'evy copula coincides with the Clayton L\'evy copula of Example \ref{claytlevycop} under the reparametrization $\theta=\frac{1}{\sigma}$.
\begin{align*}
& \mathcal{C}_{\sigma,(1,1)}
=
s_1 I\left( \frac{s_1^{-\frac{1}{\sigma}}}{s_1^{-\frac{1}{\sigma}}+s_2^{-1\frac{1}{\sigma}}},1+\sigma,1 \right)
+
s_2 I\left( \frac{s_2^{-\frac{1}{\sigma}}}{s_1^{-\frac{1}{\sigma}}+s_2^{-\frac{1}{\sigma}}},1+\sigma,1 \right)
\\
\\&=
\frac{s_1\Gamma(\sigma+2)}{\Gamma(\sigma+1)} 
\int_0^{ \frac{s_1^{-\frac{1}{\sigma}}}{s_1^{-\frac{1}{\sigma}}+s_2^{-1\frac{1}{\sigma}}} } z^\sigma\d z 
+
\frac{s_2\Gamma(\sigma+2)}{\Gamma(\sigma+1)} 
\int_0^{ \frac{s_2^{-\frac{1}{\sigma}}}{s_1^{-\frac{1}{\sigma}}+s_2^{-1\frac{1}{\sigma}}} } z^\sigma \d z 
\\&=
\frac{s_1\Gamma(\sigma+2)}{\Gamma(\sigma+1)(\sigma+1)} 
\left( \frac{s_1^{-\frac{1}{\sigma}}}{s_1^{-\frac{1}{\sigma}
}+s_2^{-1\frac{1}{\sigma}}} \right)^{\sigma+1}
+
\frac{s_2\Gamma(\sigma+2)}{\Gamma(\sigma+1)(\sigma+1)} 
\left( \frac{s_2^{-\frac{1}{\sigma}}}{s_1^{-\frac{1}{\sigma}
}+s_2^{-1\frac{1}{\sigma}}} \right)^{\sigma+1}
\\
&= \left( s_1^{-\frac{1}{\sigma	}} + s_2^{-\frac{1}{\sigma}} \right)^{-\sigma}.
\end{align*}
Hence, $\pmb{\alpha}$-Clayton L\'evy copulas contain the original Clayton L\'evy copula as a particular case and constitute an asymmetric generalization which is of interest for modelling purposes. Furthermore, this new family of L\'evy copulas retain the limit behavior of the Clayton case in $\theta$. 
\begin{theorem}\label{limbehavteo}
Let $\mathcal{C}_{\sigma,\pmb{\alpha}}$ be an $\pmb{\alpha}$-Clayton L\'evy copula and $\sigma=1/\theta$, then
$$
\lim_{\theta \to 0 } \mathcal{C}_{1/\theta,\pmb{\alpha}}
(s_1,s_2) =  \mathcal{C}_{\perp}(s_1,s_2)
$$
and
$$
\lim_{\theta \to \infty } \mathcal{C}_{1/\theta,\pmb{\alpha}}
(s_1,s_2) =  \mathcal{C}_{||}(s_1,s_2).
$$
\end{theorem}
\noindent
In Figure \ref{claycops} we show the $\pmb{\alpha}$-Clayton L\'evy copula for different choices of $\pmb{\alpha}$.

\subsection{L\'evy copulas associated to compound vectors of subordinators}
This section concludes with a result that links the survival copula associated to the score distribution of a compound vector of subordinator to the underlying L\'evy copula.  In particular, the score distribution in the compound vector of subordinators' construction has a density function $h$ which we can determine by its associated distributional survival Copula $\hat{C}$ and marginal survival functions $S_1, \ldots , S_d$.
\begin{definition}
Let $(X_1, \ldots , X_d)$ be a $d-$variate random vector and $S(x_1, \ldots , x_d)=\prob{X_1>x_1, \ldots , X_d>x_d}$ the associated $d-$variate survival function. For $i\in \left\{ 1, \ldots , d\right\}$, we say that $S_i(x)=\prob{X_i>x}$ is the $i-$th marginal survival function. The associated survival copula is given by
$$
S(x_1,\ldots , x_d)=\hat{C}
\left(
S_1(x_1),\ldots , S_d(x_d)
\right),
$$
\end{definition}
\noindent see Section 2.6 in \cite{nelsen}. The next result provides the L\'evy Copula associated to a compound vector of subordinators and provides a generalization of Theorem 5 in \cite{GL2017} where only score distributions with independent and equally distributed marginals, and hence symmetric L\'evy copulas, were considered.
\begin{theorem}\label{cormcopulateo}
Let $\pmb{Y}$ be a compound vector of subordinators given by a directing 
L\'evy measure $\nu^\star$ and a score distribution with distributional survival Copula $\hat{C}$ and marginal survival functions $S_1, \ldots , S_d$, then the L\'evy copula, $\mathcal{C}$, associated to $\pmb{Y}$ is given by
\begin{align*}
\mathcal{C}(s_1, \ldots ,  s_d)=\int_0^\infty \hat{C}
\left( S_1\left( \frac{ U_1^{-1}(s_1)  }{z} \right) , \cdots , S_d \left( \frac{ U_d^{-1}(s_d)}{z} \right) \right) \rho^\star(\d z),
\end{align*}
where the marginal tail integrals $U_i$ can be expressed as
\begin{align*}
U_i (x) = \int_{0}^\infty S_i \left( 
\frac{x}{z}
\right) \rho^\star(\d z)
\end{align*}
for $i \in \{1, \ldots d\}$.
\end{theorem}
\noindent
The above result is interesting as it shows how the dependence structure of the score distribution $h$, given by the survival Copula $\hat{C}$, and its marginal structure, given by the marginal survival functions $S_1, \ldots S_d$, impact the L\'evy copula, where the marginal structure of the vector of compound subordinators can be interpreted to be taken out with the inverse tail integrals $U_1^{-1},\ldots,U_d^{-1}$.
\section{Application to bivariate compound Poisson processes}
\noindent In this section we focus on the use of $\pmb{\alpha}$-Clayton L\'evy copulas to model bivariate compound Poisson processes. We follow \cite{esma} to perform parameter estimation. We focus on compound Poisson processes with positive increments.
\begin{definition}
Given $\lambda_1,\lambda_2 \in \re^+ \setminus \{ 0 \}$ and probability distributions $F_1$, $F_2$ in $\re^+$, a bivariate compound Poisson process with positive increments is a bivariate vector of subordinators $\left( X_1,X_2 \right)$ such that marginally $X_i$ has L\'evy intensity
\begin{align*}
\nu_i(\d s, \d x ) = \lambda_i F_i(\d s)\d x .
\end{align*}
\end{definition}
\noindent 
We observe that the associated marginal tail integrals are bounded in $\re^+$ so almost surely the associated series representation has finite jumps.
Bivariate compound Poisson processes have the form
\begin{align*}
&\left( Y_1(t), Y_2(t) \right ) \stackrel{\text{a.s.}}{=}
\\ & \left(
\sum_{i=1}^\infty W_{1,i}^\perp \indic_{\left\{ U_{1,i}^\perp \leq t  \right\}}
+
\sum_{i=1}^\infty W_{1,i}^\parallel \indic_{\left\{ U_i^\parallel \leq t  \right\}}
\ , \, 
\sum_{i=1}^\infty W_{2,i}^\perp \indic_{\left\{ U_{2,i}^\perp \leq t  \right\}}
+
\sum_{i=1}^\infty W_{2,i}^\parallel \indic_{\left\{ U_i^\parallel \leq t  \right\}}
\right),
\end{align*}
where $U_{1,i}^\perp , U_{2,i}^\perp ,W_{1,i}^\perp , W_{2,i}^\perp , U_i^\parallel,  W_{1,i}^\parallel , W_{2,i}^\parallel \stackrel{\text{a.s.}}{>}0$ for all $i\in \{1,2,\ldots \}$.
We set for $t>0$ $N_j^\perp(t)=\#\left\{i\, :\, U_{j,i}^\perp \leq t \right\}\stackrel{\text{a.s.}}{<}\infty$, with $j\in\left\{1,\right\}$, and $N^\parallel(t)=\#\left\{i\, :\, U_i^\parallel \leq t \right\}\stackrel{\text{a.s.}}{<}\infty$.
For a full review of Poisson processes we refer to \cite{kingmanpoi}. We will assume the next observation scheme for bivariate compound Poisson processes.
\begin{definition}
We say that we observe the bivariate compound process continuously over time if we are able to observe all the jump times and jump weights in a given time interval.
\end{definition}
\noindent
Let $\left\{ w_{1,i}^\perp \right\}_{i=1}^{n_1^\perp}$, $\left\{ w_{2,i}^\perp \right\}_{i=1}^{n_2^\perp}$, $\left\{ \left( w_{1,i}^\parallel, w_{2,i}^\parallel \right) \right\}_{i=1}^{n^\parallel}$ be, respectively, the jump sizes of a continuously observed bivariate compound Poisson process in a time window $[0,T]$, with $n_1^\perp = N_1^\perp(T)$ the number of jumps only appearing in dimension 1, $n_2^\perp = N_2^\perp (T)$ the number of jumps only appearing in dimension 2 and $n^\parallel =N^\parallel (T)$ the number of jumps appearing both in dimension 1 and 2.
Using the above notation we can give the likelihood for the continuous observations over time.
\begin{theorem}[\cite{esma}]\label{esmalikeli}
Let $T > 0$, if a bivariate compound Poisson process has marginal jump rates $\lambda_1$, $\lambda_2$, marginal jump weight distributions $F_i$, associated to survival functions $S_i$ and probability densities $f_i$ parameterized by real valued vectors $\pmb{c}_j$, $j\in \{1, 2\}$, and an associated L\'evy copula $\mathcal{C}_{\pmb{k}}$ parameterized by a real valued vector $\pmb{k}$ such that $\frac{\partial^2}{\partial u_1  \partial u_2}\mathcal{C}_{\pmb{k}}(u_1,u_2)$ exists for every $(u_1,u_2,x)\in (0,\lambda_1 ) \times (0, \lambda_2 )\times \re^+$; then the likelihood function for continuously observed bivariate compound Poisson processes over $(0,T]$ is given by
\begin{align*}
L(\lambda_1, & \lambda_2, \pmb{c}_1,  \pmb{c}_2, \pmb{k})
= 
(\lambda_1)^{n_1^\perp} e^{-\lambda_1^\perp T}
\prod_{i=1}^{n_1^\perp}
\left(
f_1( w_{1,i}^{n_1^\perp}; \pmb{c}_1 )
\left( 1 - \restr{ \frac{\partial}{\partial u_1 } \mathcal{C}_{\pmb{k}} (u_1,\lambda_2) }{ u_1 = \lambda_1 S_1(w_{1,i}^\perp ; \pmb{c}_1) }
\right)
\right)
\\
&\times 
(\lambda_2)^{n_2^\perp} e^{-\lambda_2^\perp T}
\prod_{i=1}^{n_2^\perp}
\left(
f_2( w_{2,i}^{n_1^\perp}; \pmb{c}_2 )
\left( 1 - \restr{ \frac{\partial}{\partial u_2 } \mathcal{C}_{\pmb{k}} (\lambda_1,u_2) }{ u_2 = \lambda_2 S_2(w_{2,i}^\perp ; \pmb{c}_2) }
\right)
\right)
\\
&\times
( \lambda_1 \lambda_2 )^{n^\parallel }
e^{-\lambda^\parallel T} 
\prod_{i=1}^{n^\parallel}
\left(
f_1( w_{1,i}^\parallel; \pmb{c}_1)
f_2( w_{2,i}^\parallel; \pmb{c}_2)
\right.
\\
& \left.
\times
\restr{
\frac{\partial^2}{\partial u_1 \partial u_2}\mathcal{C}_{\pmb{k}}
(u_1,u_2)
}{ 
u_1 = \lambda S_1( w_{1,i}^\parallel ; \pmb{c}_1 )
, \, 
u_2 = \lambda S_2( w_{2,i}^\parallel ; \pmb{c}_2 ) 
}
\right),
\end{align*}
with $\lambda^\parallel = \mathcal{C}_{\pmb{k}}(\lambda_1,\lambda_2)$ and $\lambda_j^\perp = \lambda_j - \lambda^\parallel $ for $j\in \left\{ 1,2 \right\}$.
\end{theorem}
\noindent
The application of our extension of the Clayton L\'evy copula $\mathcal{C}_{\sigma,\pmb{\alpha}}$ is of interest for the above model as it can offer more flexibility in the above likelihood.  
\subsection{Working example simulation study}
We will use the above likelihood to perform maximum likelihood estimation for our working example compound vector of subordinators in Theorem \ref{workexpteo}.
As discussed in \cite{esma2}, we can assume an observation scheme for vectors of subordinators where only jump weights greater than some thresholds $\epsilon_i$ are observed in the $i$-th dimension of the vector. Let $\rho$ be a bivariate L\'evy intensity with tail integral $U$ and marginal tail integrals $U_i$. Let $\pmb{\epsilon} =(\epsilon_1, \ldots, \epsilon_d)$ have positive non-zero entries, $\lambda_i^{(\epsilon_i)}=U_i(\epsilon_i)$ and set $\rho_i(\d s) = \lambda_i
F_i(\d s)$ and $\lambda^{(\pmb{\epsilon})}=U(\epsilon_1, \ldots , \epsilon_d)$. If $n^{\pmb{\epsilon}}$ is the number of observations with jump weights $(w_{1,i},w_{2,i})$ attaining the thresholds as discussed above, the likelihood in Theorem \ref{esmalikeli} is given as follows:
\begin{theorem}[\textbf{\cite{esma2}}]
Let $T>0$, if $\rho_i$ is parameterized by real valued vectors $\pmb{c}_i$, $i\in \{1, 2\}$, and an associated L\'evy copula $\mathcal{C}_{\pmb{k}}$ parameterized by a real valued vector $\pmb{k}$ such that $\frac{\partial^2}{\partial u_1  \partial u_2}\mathcal{C}_{\pmb{k}}(u_1,u_2)$ exists for every $(u_1,u_2,x)\in (0,\lambda_1 ) \times (0, \lambda_2 )\times \re^+$, then the likelihood for the observed vector of subordinators with jump weights above thresholds $\pmb{\epsilon}$ is given by
\begin{align*}
&L^{(\pmb{\epsilon})}(\pmb{c}_1,\pmb{c}_2,\pmb{k})=
\\
&e^{-\lambda^{(\pmb{\epsilon})}T}
\prod_{i=1}^{n^{\pmb{\epsilon}}}
\rho_1(w_{1,i} ; \pmb{c}_i )
\rho_2(w_{2,i} ; \pmb{c}_2 )
\restr{
\frac{\partial^2}{\partial u_1 \partial u_2}\mathcal{C}_{\pmb{k}}
(u_1,u_2)
}{ 
u_1 = U_1( w_{1,i} ; \pmb{c}_1 )
, \, 
u_2 = U_2( w_{2,i} ; \pmb{c}_2 ) 
}.
\end{align*} 
\end{theorem}
\noindent
To draw observations from our working example as described above we use the series representation \eqref{compserrepr} with a threshold of $\tau=10^{-8}$ and fix $\alpha_1=1$, $\beta_1=2$, $\alpha_2=10$, $\beta_2=5$, $\theta=0.5$ and $K=1$ in \eqref{workexintens}. For the likelihood in the above theorem we choose thresholds of $\epsilon_1 = 10^{-6}$ and $\epsilon_2 = 10^{-5}$. We use the Nelder-Mead algorithm, see \cite{nelder} and \cite{gao}, from the Optim.jl Julia package, \cite{optim}, to numerically optimize the above likelihood for $\alpha_1,\beta_1,\alpha_2,\beta_2$ and $\sigma$, we fix $K=1$ in order to avoid identifiability issues with the parameters $\beta_1$ and $\beta_2$. In table \ref{tablesim} we show fitted values for simulation studies with $T=1$ and $T=100$.

\begin{table}[h!]
\begin{center}
\begin{tabular}{r|cccc}
\toprule
Parameter  & True value & Fitted value with $T=1$     & Fitted value with $T=100$ \\ \hline
$\alpha_1$ & 1.0 &  1.037 & 1.003 \\
$\beta_1$ & 2.0 &   2.282 & 2.036 \\
$\alpha_2$ & 10.0 &  8.726 & 10.231\\
$\beta_2$ & 5.0 &    4.909 & 5.151 \\
$\sigma$ & 0.5 &     0.509 & 0.500\\
\bottomrule
\end{tabular}
\end{center}
\caption{Maximum likelihood fits for working example compound vector of subordinators.}
\label{tablesim}
\end{table}
\noindent

\begin{figure}[h!]
\begin{center}
\includegraphics[width=0.5\linewidth]{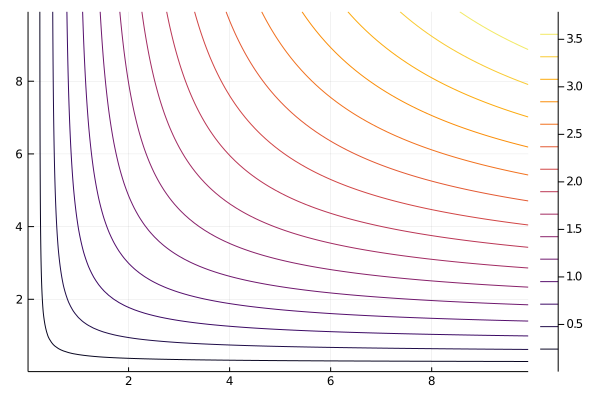}
\includegraphics[width=0.5\linewidth]{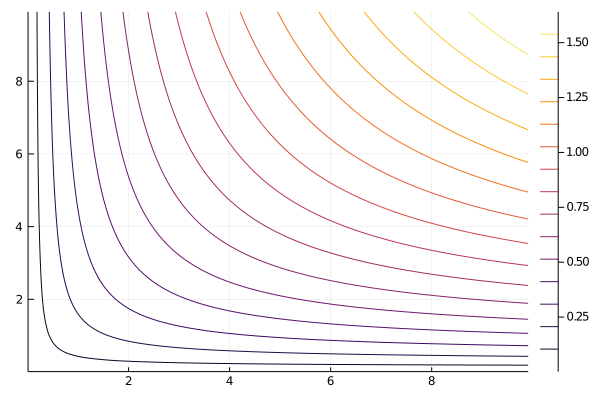}
\includegraphics[width=0.5\linewidth]{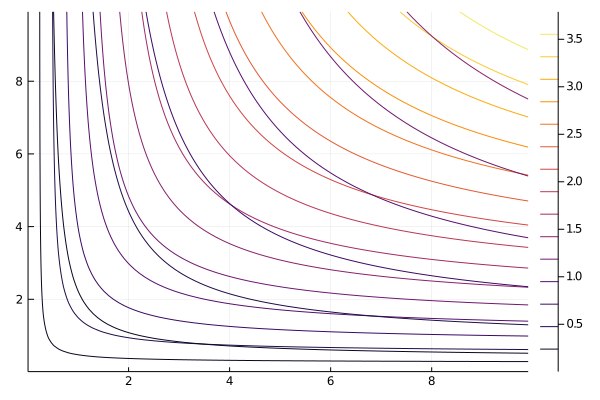}
\caption{Top: True $(1,10)$-Clayton L\'evy copula. Middle: Miss-specified fitted symmetric $(0.355,0.355)$-Clayton L\'evy copula. Bottom: crossing of both forementioned $\pmb{\alpha}$-Clayton L\'evy copulas, darker lines correspond to the symmetric case as can be seen from the contour levels of the previous plots.}\label{claycops}
\end{center}
\end{figure}
\clearpage
\noindent
We also fitted a miss-specified symmetric model with fixed $\alpha_1=\alpha_2=\alpha$ and $\beta_1=\beta_2=1$, such model coincides with the use of a Clayton L\'evy copula when $\alpha=1$. For $T=100$ we obtained fits $\hat{\alpha}=0.355$ and $\hat{\sigma}=0.625$. For illustration purpose we show in Figure \ref{claycops} the fitted L\'evy copulas for the miss-specified and specified model, where we can appreciate the inadequacies of modeling asymmetric observations with a symmetric L\'evy copula.
\subsection{Danish fire insurance dataset study}
\begin{figure}[h]
\begin{center}
\includegraphics[width=0.7\linewidth]{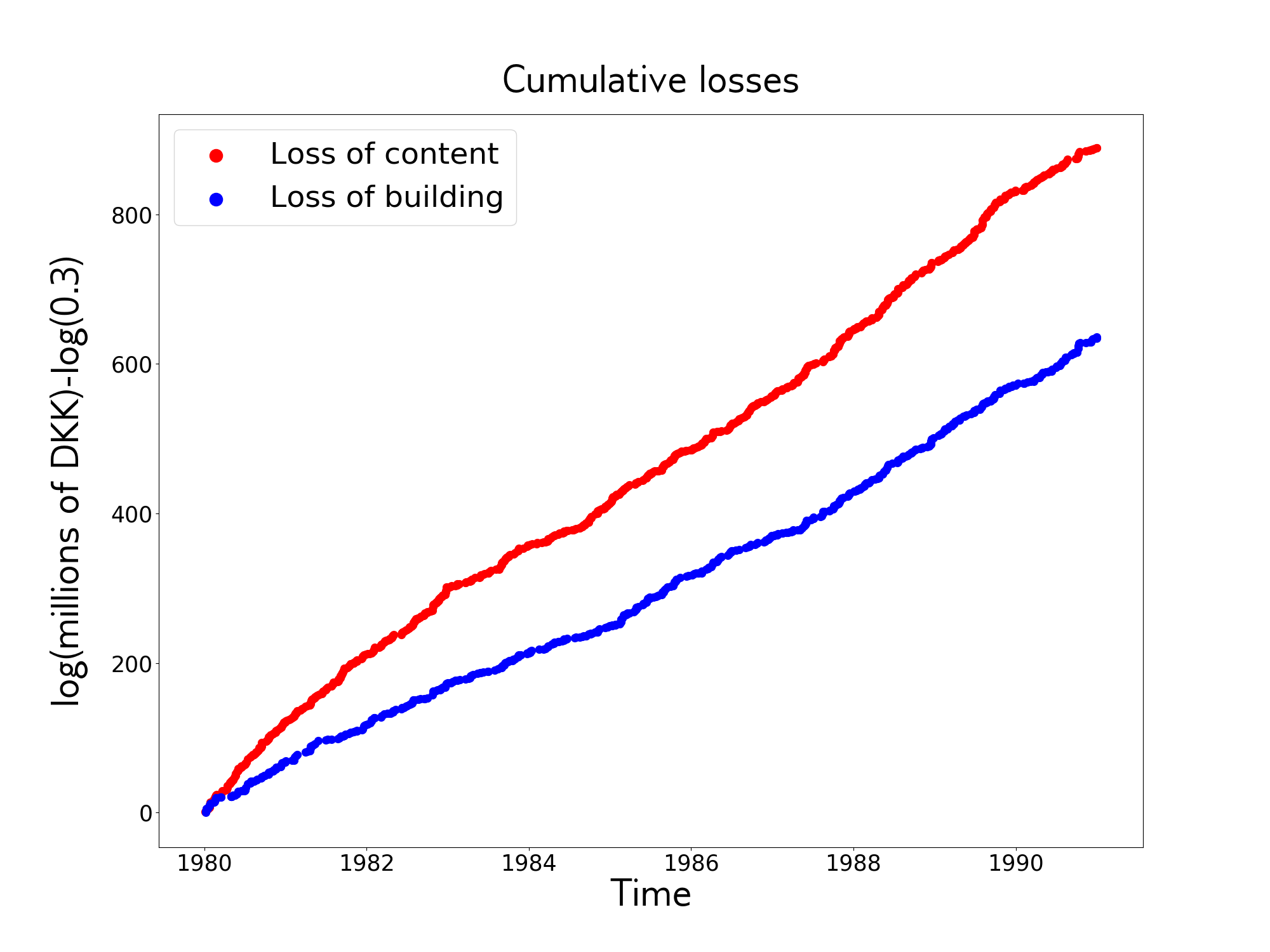}
\caption{Cumulative logarithmic losses related to loss of content (blue) and loss of building (red) in the Danish fire insurance data as discussed in Section 5.}\label{fig:test1}
\end{center}
\end{figure}

\begin{figure}
\begin{center}
\includegraphics[width=0.7\linewidth]{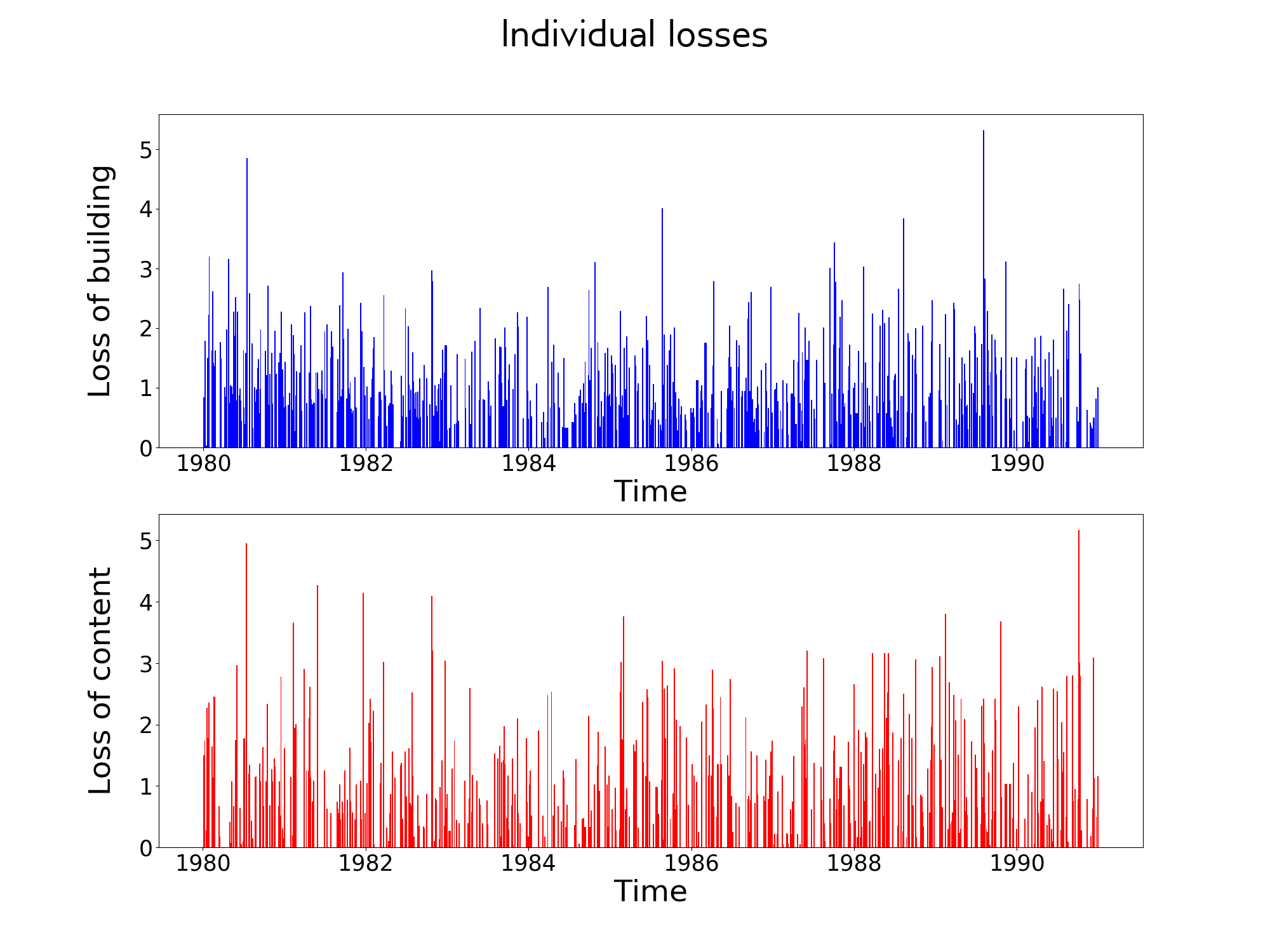}
\caption{Individual logarithmic losses related to losses due to the building (top) and losse due to content (bottom) in the Danish fire insurance data as discussed in Section 5.}  \label{fig:test2}
\end{center}
\end{figure}

\begin{figure}
\begin{center}
Marginal fits
\\
\includegraphics[width=0.48\linewidth]{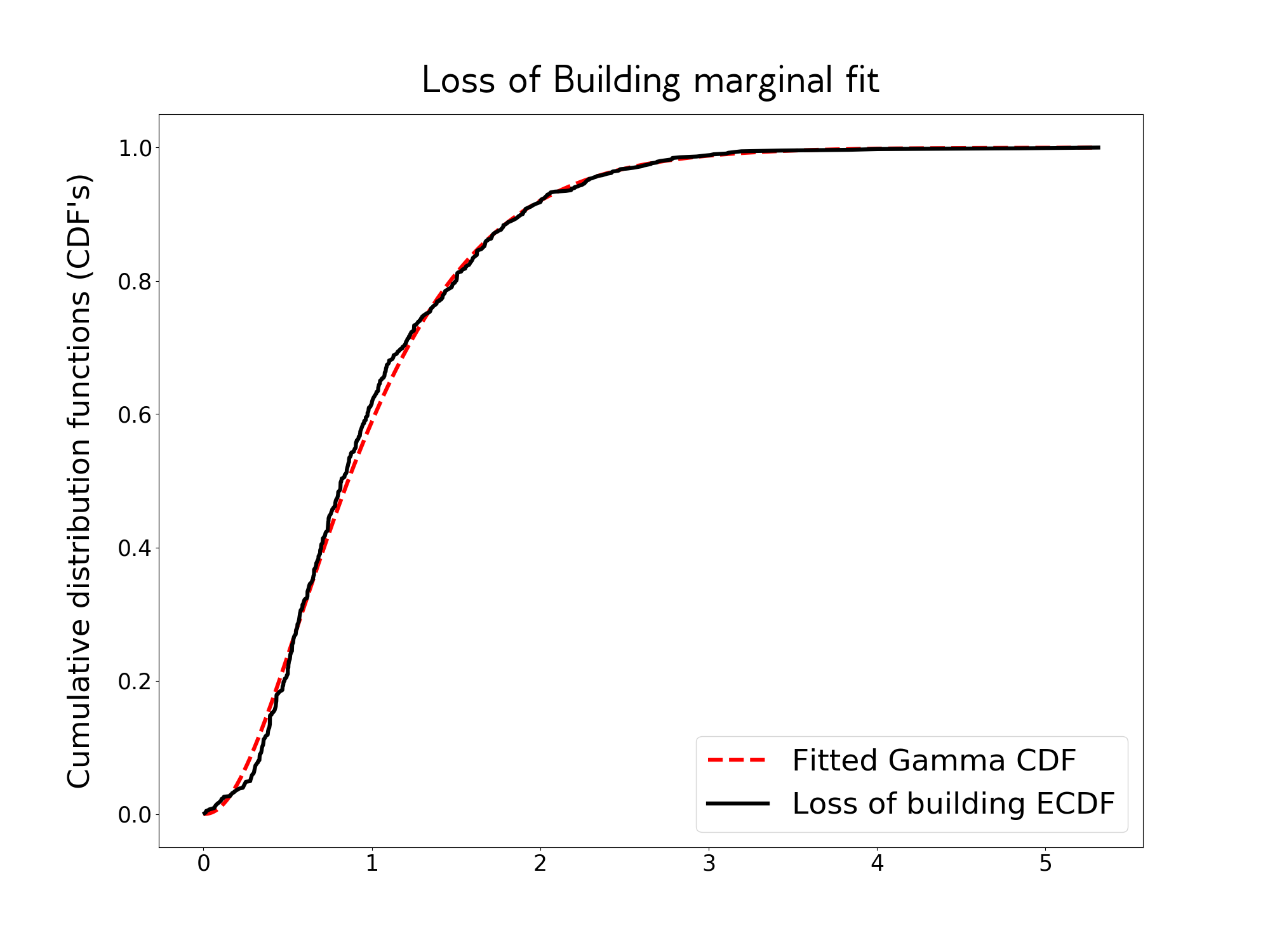}
\includegraphics[width=0.48\linewidth]{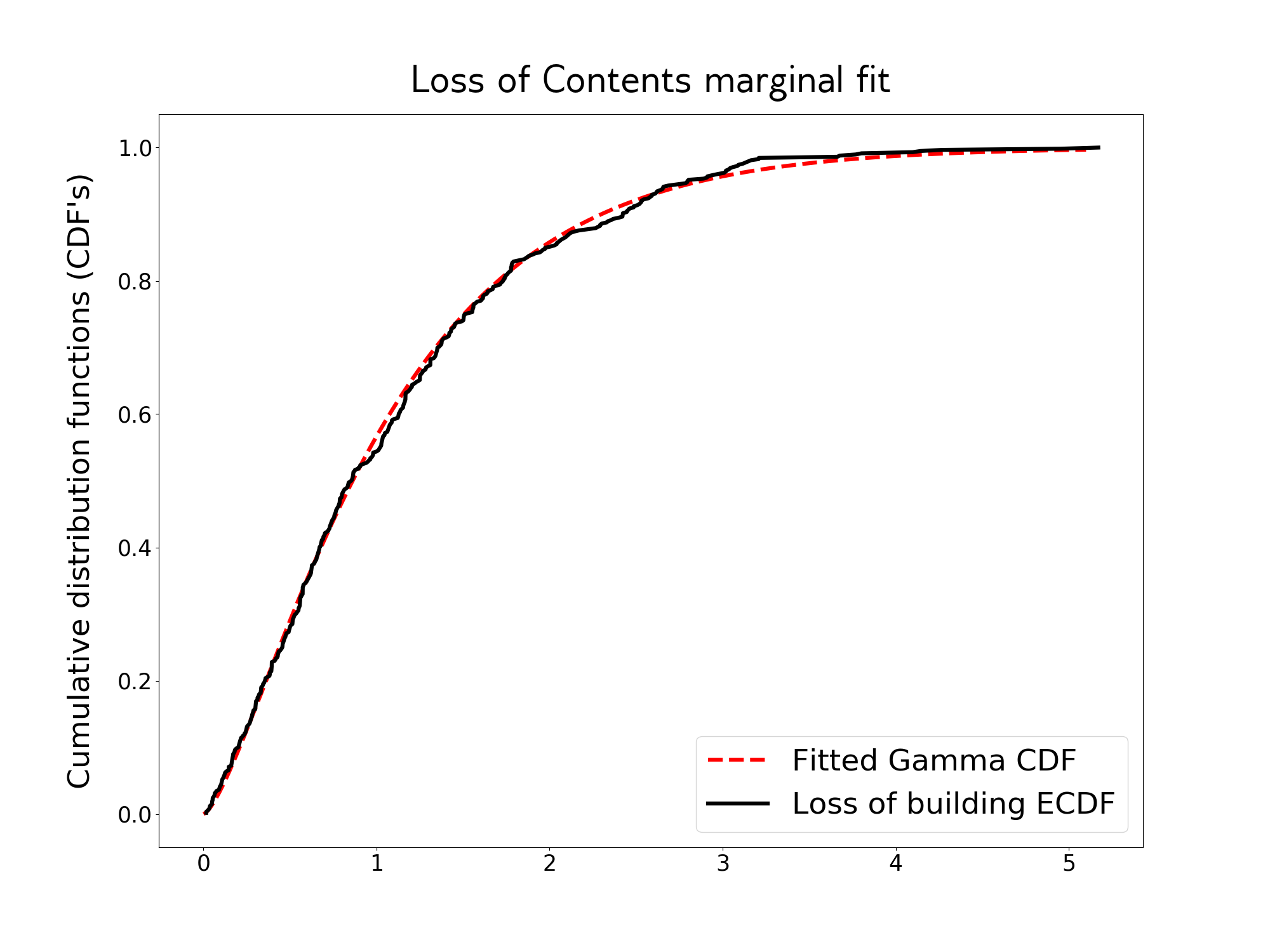}
\caption{Marginal fit for the marginal cumulative distribution function (CDF) associated to the losses due to the building (top) and due to content (bottom) in the Danish fire insurance dataset as discussed in Section 5. The black lines correspond to the empirical cumulative distribution function (ECDF) and the red lines corresponds to the maximum likelihood fits with Gamma distributions.}\label{cdf_marg_fits}
\end{center}
\end{figure}

\begin{figure}
\begin{center}
Independent losses fits
\\
\includegraphics[width=0.48\linewidth]{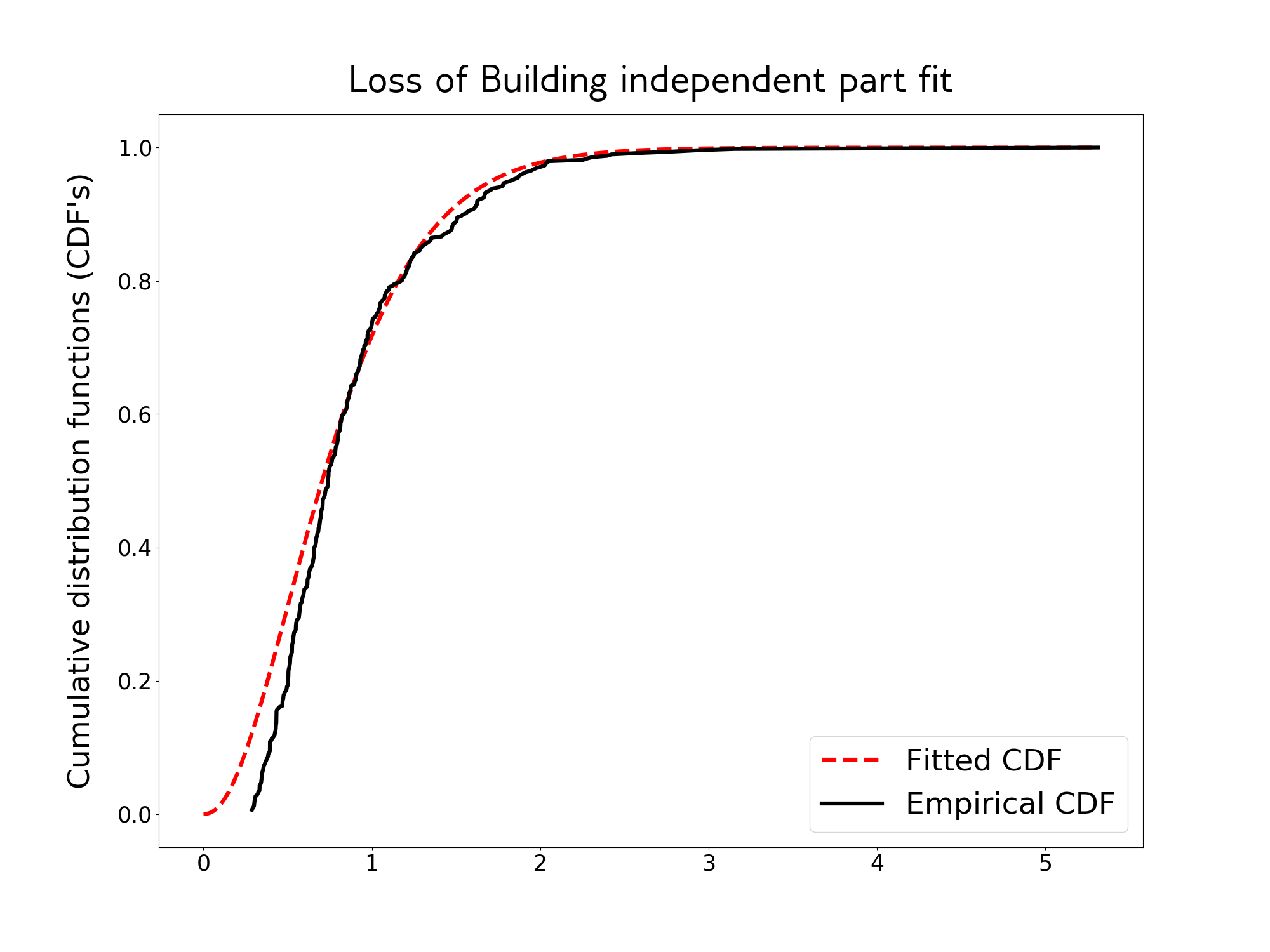}
\includegraphics[width=0.48\linewidth]{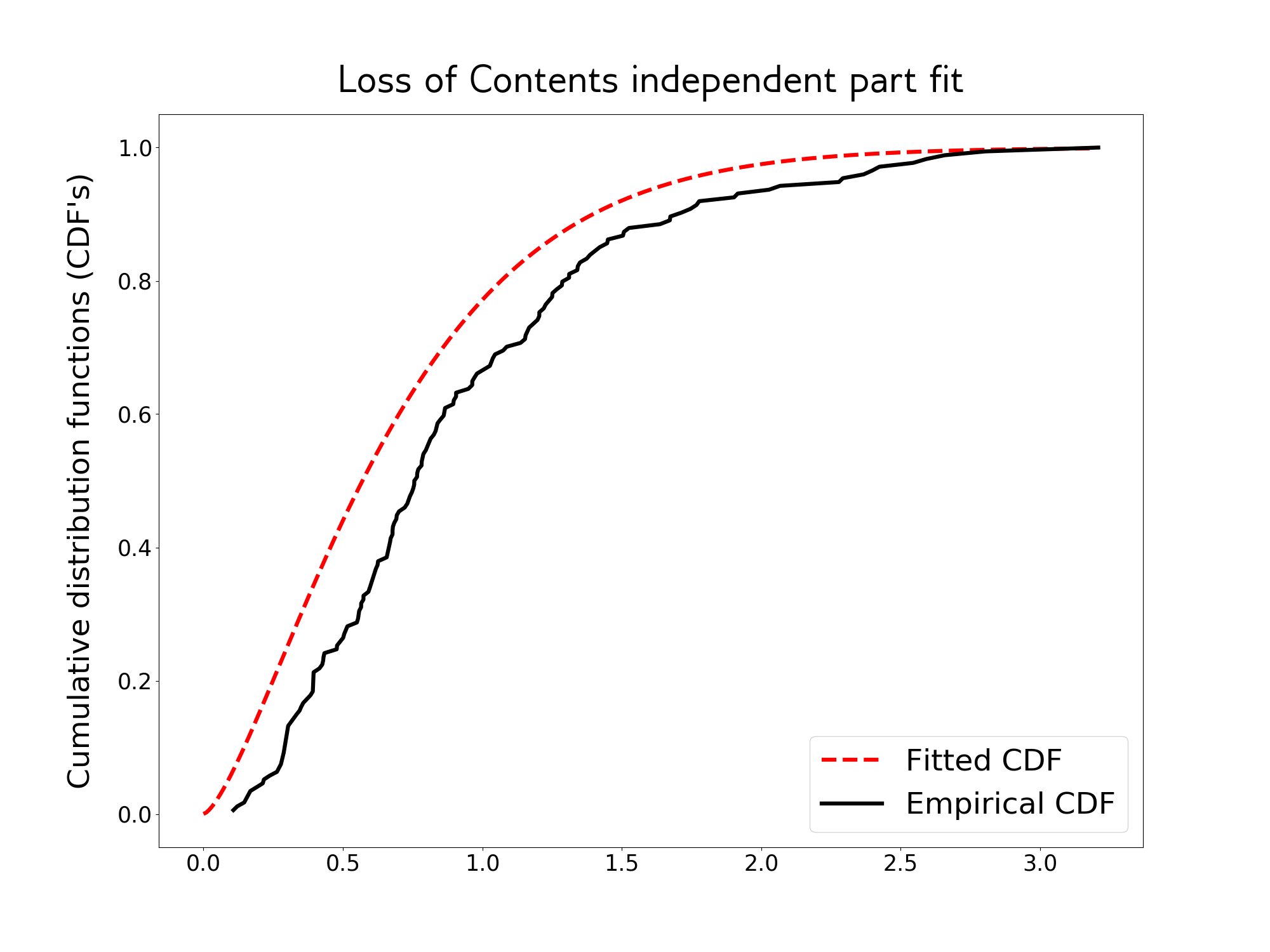}
\caption{Marginal fit for the independent losses, $w_{1,i}^\perp$, $w_{2,i}^\perp$, cumulative distribution function (CDF) associated to the losses due to the building (top) and due to content (bottom) in the Danish fire insurance dataset as discussed in Section 5. The black lines correspond to the empirical cumulative distribution function (ECDF) and the red lines corresponds to the maximum likelihood fits for the $\pmb{\alpha}$-Clayton L\'evy copula.}\label{indep_fits}
\end{center}
\end{figure}

\begin{figure}
\begin{center}
Dependent losses fits
\\
\includegraphics[width=0.48\linewidth]{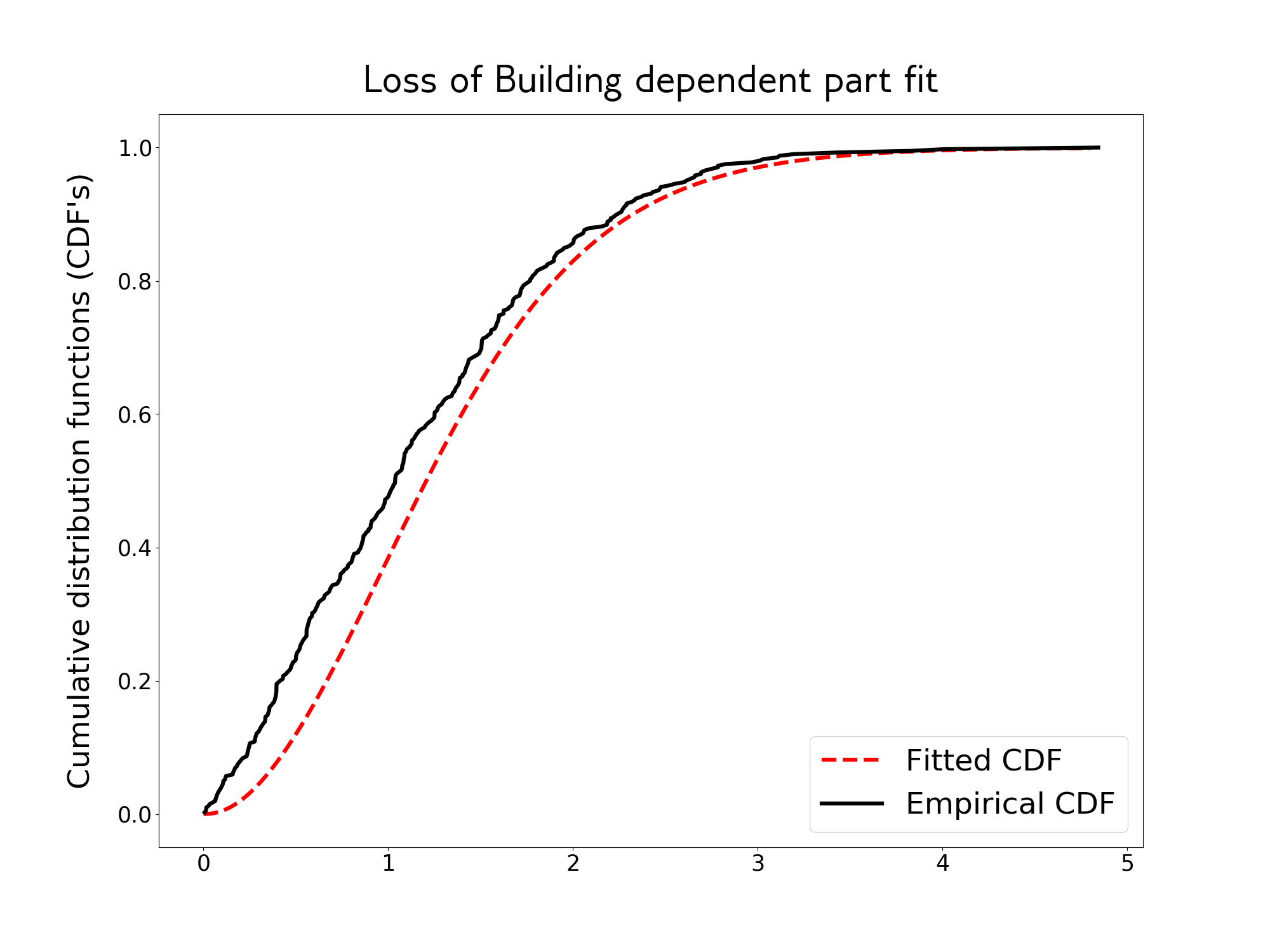}
\includegraphics[width=0.48\linewidth]{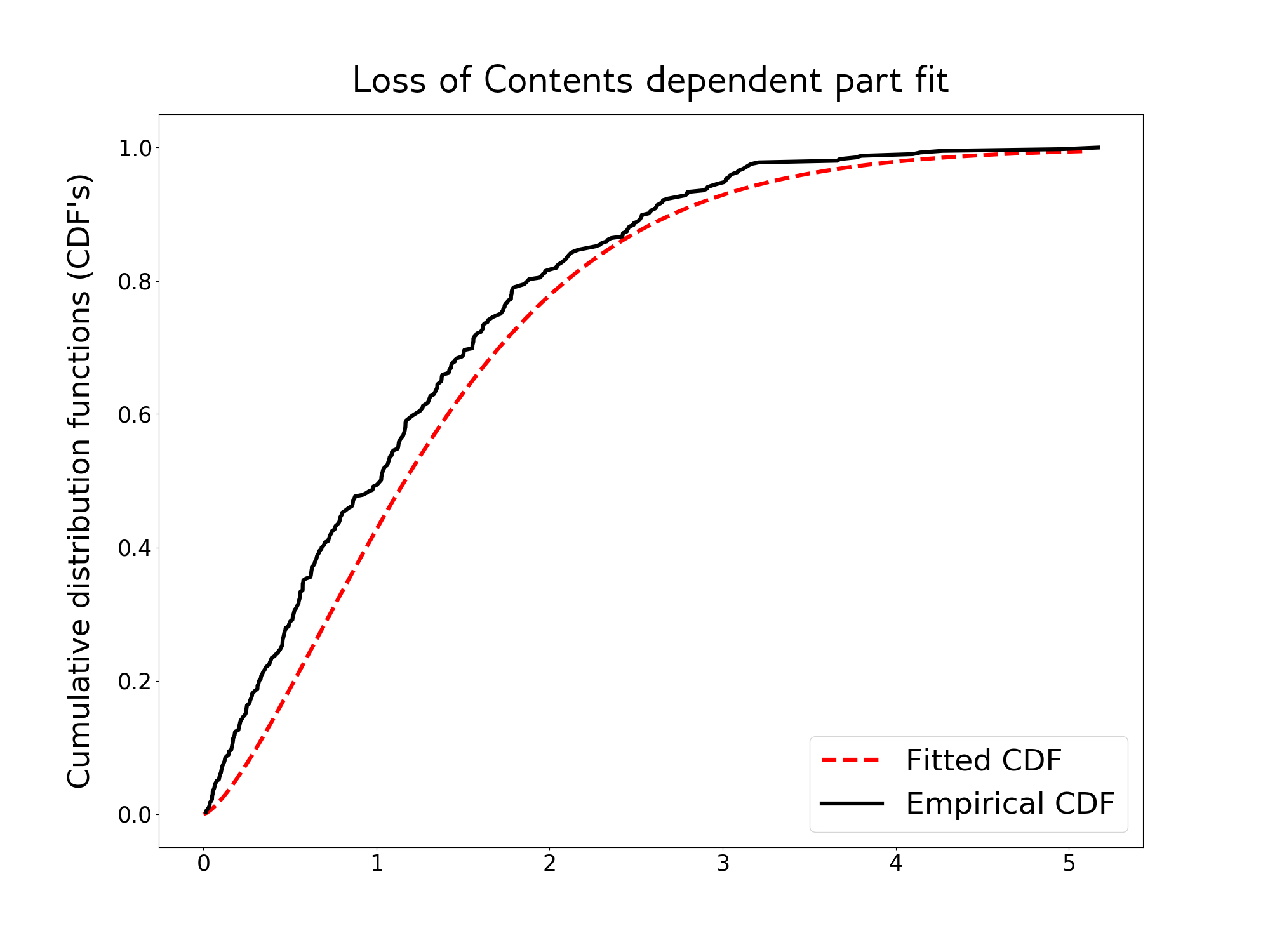}
\caption{Marginal fit for the dependent losses, $(w_{1,i}^\parallel , w_{2,i}^\parallel )$, cumulative distribution function (CDF) associated to the losses due to the building (top) and due to content (bottom) in the Danish fire insurance dataset as discussed in Section 5. The black lines correspond to the empirical cumulative distribution function (ECDF) and the red lines corresponds to the maximum likelihood fits for the $\pmb{\alpha}$-Clayton L\'evy copula.}\label{dep_fits}
\end{center}
\end{figure}

In tis section we perform a real data analysis of the Danish fire insurance dataset available in the "fitdistrplus" R package, see \cite{fitdistrplus}. The data consist of the losses, in millions of Danish Krone, pertaining to 2167 fire incidents in Copenhagen between 1980 to 1990.
We follow  \cite{esma} and take only into account the building and content losses. Furthermore we focus on losses such that the loss due to the building and due to the contents are both greater than 750000 Danish Kronen or one is greater than 750000 Danish Kronen while the other is zero, so we get 1066 observations. Following their approach, we consider the logarithm of the loss quantities and normalize them by subtracting $\log(0.75)$; thus obtaining the bivariate weights, at each time point, which we model through a bivariate compound Poisson process. In Figures \ref{fig:test1} and \ref{fig:test2} we show the corresponding bivariate Poisson process. We fit the model in a two-step way by following \cite{esma3}, see also \cite{jiang}.  In particular, we fit the marginal parameters first and the dependence parameters in the L\'evy copula secondly. The marginal distributions, $F_1$ and $F_2$ are fitted via maximum likelihood and modeled  with Gamma distributions, as in comparison with Weibull and LogNormal distributions such choice gave a lower uniform distance between the fitted cumulative distribution function and the empirical; this marginal fits are showed in Figure \ref{cdf_marg_fits}. We fit the $\pmb{\alpha}$-Clayton L\'evy copula parameters $\theta=\frac{1}{\sigma}$, $\alpha_1$ and $\alpha_2$ using maximum likelihood. In Figures \ref{indep_fits} and \ref{dep_fits} we show the fitted cumulative distributions for the dependent component losses of building and contents, $(w_{1,i}^\parallel, w_{2,i}^\parallel)$, and the independent component losses due to building and contents, $w_{1,i}^\perp$, $w_{2,i}^\perp$, with the respective empirical cumulative distributions for comparison. We observe that the model shows flexibility in fitting the dependent and independent components while the marginal fits as in Figure \ref{cdf_marg_fits} are kept fixed. We used again the Nelder-Mead algorithm, see previous subsection, and found a maximum loglikelihood value of $-4920.20$. We also found the maximum likelihood estimators for the $\pmb{\alpha}$-Clayton L\'evy copula constrained to be symmetric, $\alpha_1=\alpha_2$ and for the Clayton L\'evy copula, $\alpha_1=\alpha_2=1$, where we got maximum loglikelihood values of, respectively $-4925.62$ and $-4925.75$. It is natural that the unrestricted model can attain a higher likelihood value, which for datasets presenting asymmetry on the underlying L\'evy copula must have a significative difference.



\clearpage

\section{Discussion}

Diverse families of L\'evy copulas have been proposed such as \textit{archimedean} L\'evy copulas, see Proposition 5.6 and 5.7 in \cite{tankbook}, \textit{vine} L\'evy copulas, see \cite{grothe} and \textit{pareto} L\'evy copulas, see \cite{eder}. Pareto and archimedean L\'evy copulas are symmetric by construction while vine L\'evy copulas can be asymmetric either by using an asymmetric distributional or L\'evy copula in their construction. $\pmb{\alpha}$-Clayton L\'evy copulas thus help to enrich the examples of asymmetric L\'evy copulas in the literature. Extension to arbitrary dimension $d$ is possible by generalizing Theorem \ref{workexpteo} into a $d$-variate setting by using a $d$-variate score distribution which has independent Gamma marginals, however the change of variable for obtaining the tail integral, see the proof of Theorem \ref{claylevycopteo}, becomes analytically cumbersome as it involves the cumulative distribution function of a Dirichlet distribution. Further choices for the score distribution and directing L\'evy measure of a compound vector of subordinators can be considered. An example that we have found to be useful in applications is to use LogNormal score distributions with which the mass of the score distribution can be adequately distributed distributed in $d$-dimensions, see \eqref{serrepr}. Such choice of score distribution is seen to define a well posed compound vector of subordinators by use of Theorem \ref{integcondteo}. However such choice seems to usually not be analytically tractable, for example with $\sigma$-stable and gamma directing L\'evy measures. The study of the numerical treatment for the L\'evy copulas associated to compound vectors of subordinators is left as future work.

\clearpage
\section*{Acknowledgments}
Fabrizio Leisen was supported by the European Community's Seventh Framework Programme [FP7/2007-2013] under grant agreement no: 630677.

\appendix

\section{Proofs}

\noindent\subsection*{Proof of Theorem \ref{workexpteo} }

By Definition 4, the corresponding L\'evy intensity is
\begin{align*}
\rho(\d s_1, \d s_2)&=
\frac{\sigma K \beta_1^{\alpha_1}\beta_2^{\alpha_2}
s_1^{\alpha_1 -1}s_2^{\alpha_2 - 1}
}{\Gamma (\alpha_1)\Gamma (\alpha_2)}
\int_0^\infty
\left(\frac{1}{z}\right)^{2+\alpha_1 +\alpha_2+\sigma -1 }
e^{-(\beta_1 s_1 +\beta_2 s_2)\frac{1}{z}}
\d z
\\&=
\frac{\sigma K \beta_1^{\alpha_1}\beta_2^{\alpha_2}
s_1^{\alpha_1 -1}s_2^{\alpha_2 - 1}
}{\Gamma (\alpha_1)\Gamma (\alpha_2)}
\int_0^\infty
u^{\alpha_1 +\alpha_2+\sigma -1 }
e^{-(\beta_1 s_1 +\beta_2 s_2)u}
\d u
\\&=
\frac{\sigma K \beta_1^{\alpha_1}\beta_2^{\alpha_2}
\Gamma(\alpha_1 +\alpha_2 +\sigma)s_1^{\alpha_1 -1}s_2^{\alpha_2 - 1}
}{\Gamma (\alpha_1)\Gamma (\alpha_2)(\beta_1 s_1 +\beta_2 s_2)^{\alpha_1+\alpha_2+\sigma}}
\end{align*}
And the corresponding marginals are given by
\begin{align*}
\rho_i(\d s)& =
\frac{\sigma K \beta_i^{\alpha_i}
s^{\alpha_i -1}
}{\Gamma (\alpha_i)}
\int_0^\infty
\left(\frac{1}{z}\right)^{2+\alpha_i+\sigma -1 }
e^{-\beta_i s\frac{1}{z}}
\d z
\\
&=\frac{\sigma K \beta_i^{\alpha_i}
s^{\alpha_i -1}
}{\Gamma (\alpha_i)}
\int_0^\infty
u^{\alpha_i+\sigma -1 }
e^{-\beta_i s u}
\d u
\\
&=\frac{\sigma K \beta_i^{-\sigma} 
\Gamma(\alpha_i+\sigma) s^{-\sigma -1}
}{\Gamma (\alpha_i)}.
\end{align*}
The marginal tail integral is
\begin{align*}
U_i(y)=\int_y^\infty
\rho_i(\d s)
= \frac{ K 
\beta_i^{-\sigma}\Gamma(\alpha_i+\sigma)
}{\Gamma (\alpha_i)} \int_y^\infty \sigma s^{-\sigma -1}
= \frac{ K 
\beta_i^{-\sigma}\Gamma(\alpha_i+\sigma)
y^{-\sigma}
}{\Gamma (\alpha_i)}
\end{align*}
which has inverse
\begin{align*}
U_i^{-1}(y)= \left( \frac{\Gamma(\alpha_i)y}{K\beta_i^{-\sigma}\Gamma(\alpha_i+\sigma)} \right)^{-\frac{1}{\sigma}}
\end{align*}

\noindent \subsection*{Proof of Theorem \ref{serreprteo} }

\noindent
For this proof we will use Proposition 2.1 in \cite{rosinski}. Let $H$ be the probability distribution associated to $h$ and $\nu^\star$ the directing L\'evy intensity.
We consider a Poisson random measure
$$
M = \sum_{i=1}^\infty \delta_{
\left(
\pmb{Z}_i, W_i, X_i
\right)},
$$
where $\{\pmb{Z}_i\}_{i=1}^\infty \stackrel{\text{i.i.d.}}{\sim}H$ and $\{ \left( W_i, X_i \right) \}_{i=1}^\infty$ are such that
$$
\sum_{i=1}^\infty \delta_{
\left(
 W_i, X_i
\right)},
$$ as a Poisson random measure, has intensity $\nu^\star$. It follows that $M$ has intensity $\mu=H \times \nu^\star$. 
We define
$$
g(\pmb{z}, w, x)=(z_1 w, z_2 w, \ldots , z_d w , x).
$$
Due to Proposition 2.1 in \cite{rosinski} it suffices to check that
$\nu = \mu \circ g^{-1} $.
Let $ A_1, \ldots , A_d ,B \in  \mathcal{B} \left( \re^+ \right)$, then
\begin{align*}
&g^{-1}\left( (A_1\times \ldots \times A_d ) \times B \right)=
\\
&
\left\{ \left( \frac{a_1}{w}, \frac{a_2}{w}, \ldots , \frac{a_d}{w}, w, x\right) \text{ such that } x \in B, \; a_1 \in A_1, \ldots a_d \in A_d, \; w \in \re^+ \right\}
\end{align*}
So the pullback measure $\eta = \mu\circ g^{-1}$ is given by
\begin{align*}
&\eta \left( (A_1 \times \ldots \times \times A_d) \times B \right) =\int_{g^{-1}( (A_1 \times \ldots \times \times A_d) \times B  )} \mathrm{d}\mu 
\\
&
= \int_{ A_1/z \times A_2/z \times \ldots \times A_d/z \times (0,\infty)\times B } H( \d s_1, \ldots , \d s_d) \nu^\star(\d z, \d x )
\\
&
=
\int_{ A_1 \times A_2 \times \ldots \times A_d \times (0,\infty)\times B } H\left( \frac{\d s_1}{z}, \ldots , \frac{\d s_d}{z} \right) \nu^\star(\d z, \d x )
\end{align*}
So extending the measure we conclude that $\nu=\mu \circ g^{-1}$ so
$$
N=\sum_{i=1}^\infty \delta_{(Z_{1,i}W_i, Z_{2,i}W_i, \ldots , Z_{d,i}W_i, X_i )}
$$
is almost surely a compound vector of subordinators given by the score distribution $h$ and the directing L\'evy measure $\nu^\star$ due to Proposition 2.1 in \cite{rosinski}.

\noindent\subsection*{Proof of Theorem \ref{integcondteo} }
\noindent
Let $|\pmb{w}|=\sum_{i=1}^d w_i$ for $\pmb{w}\in (\re^+)^d$. We have that $\| z\pmb{w}\|\leq z|\pmb{w}|$ so
\begin{align*}
&\int_{(\re^+)^d \times \re^+ }
\min \left\{
1, \| \pmb{s} \|
\right\}
\nu(\d \pmb{s}, \d x )
= 
\int_{(\re^+)^d \times \re^+ \times \re^+ }
\min \left\{
1, \| z\pmb{w} \|
\right\}
h( \pmb{w} )\d \pmb{w}  \nu^\star (\d z, \d x )
\\ & \leq 
\int_{(\re^+)^d \times \re^+ \times \re^+ }
\min \left\{
1, z|\pmb{w}|
\right\}
h(\pmb{w})\d \pmb{w}  \nu^\star (\d z, \d x )
=
\esp{
\int_{ (\re^+)^2 }
\min \left\{
1, z |\pmb{W} |
\right\}
\nu^\star (\d z, \d x )
}
\\&=
\esp{
\int_{ \left(0,\frac{1}{|\pmb{W}|} \right)\times\re^+ }
z |\pmb{W} |
\nu^\star (\d z, \d x )
} + \esp{ 
\int_{  \left[ \frac{1}{|\pmb{W}|},\infty \right)\times \re^+ }
\nu^\star (\d z, \d x )
}
\\
&=
\esp{
\int_{ \re^+ \times\re^+ }
\indic_{\left\{
z < \frac{1}{|\pmb{W}|}
\right\}}
z |\pmb{W} |
\nu^\star (\d z, \d x )
} + \esp{ 
\int_{  \re^+ \times \re^+ }
\indic_{\left\{
\frac{1}{|\pmb{W}|} \leq z
\right\}}
\nu^\star (\d z, \d x )
}
\\&=
\int_{ (0,1)\times\re^+ }
\esp{
\indic_{\left\{ |\pmb{W}| < \frac{1}{z} 
\right\}}
|\pmb{W}|}z
\nu^\star (\d z, \d x )
+
\int_{ [1,\infty) \times \re^+ }
\esp{
\indic_{\left\{ |\pmb{W}| < \frac{1}{z} 
\right\}}
|\pmb{W}|}z
\nu^\star (\d z, \d x )
\\&\hphantom{=}+
\int_{ (0,1)\times\re^+ }
\esp{
\indic_{\left\{ |\pmb{W}| \geq \frac{1}{z} 
\right\}}
}
\nu^\star (\d z, \d x )
+
\int_{ [1,\infty) \times \re^+ }
\esp{
\indic_{\left\{ |\pmb{W}| \geq \frac{1}{z} 
\right\}}
}
\nu^\star (\d z, \d x )
\\&\leq
\esp{|\pmb{W}|}
\int_{ (0,1)\times\re^+ }z
\nu^\star (\d z, \d x )
+
\int_{ [1,\infty) \times \re^+ }
\nu^\star (\d z, \d x )
\\&\hphantom{=}+
\esp{|\pmb{W}|}
\int_{ (0,1)\times\re^+ }
z
\nu^\star (\d z, \d x )
+
\int_{ [1,\infty) \times \re^+ }
\nu^\star (\d z, \d x )<\infty.
\end{align*}
For the first and fourth integral we use the fact that the indicator function is less or equal to one. For the second integral, we note that $\esp{
\indic_{\left\{ |\pmb{W}| < \frac{1}{z} 
\right\}}
z|\pmb{W}|}\leq 1$. For the third integral, we use the Markov's inequality, $\esp{
\indic_{\left\{ |\pmb{W}| \geq \frac{1}{z} 
\right\}}
}=\prob{|\pmb{W}|\geq \frac{1}{z} } \leq z\esp{|\pmb{W}|}$. Finiteness of the above expression follows  from the fact that $\nu^\star$ is a L\'evy intensity satisfying \eqref{integcond} and $\esp{|\pmb{W}|}<\infty$.
\noindent\subsection*{Proof of Theorem \ref{compoundmomentsteo} }
We can set the multivariate L\'evy intensity for a $d-$variate vector of subordinators $\pmb{Y}=\left( Y_1, \ldots , Y_d \right)$ with $\pmb{\lambda}=\left( \lambda_1, \ldots , \lambda_d \right)$ as
$$
\psi_t (\pmb{\lambda})=\int_0^t \int_0^\infty (1-e^{-\lambda_1  s_1 - \ldots - \lambda_d s_d})\nu(\d s, \d x)=-\log\left( \esp{e^{-\lambda_1 Y_1(t)-\ldots - \lambda_d Y_d  } } \right).
$$
If we denote $\{\pmb{e}_i\}_{i=1}^d$ as the canonical basis of $\re^d$ and the univariate Laplace exponent of $Y_i$ as $\psi_t(\lambda \pmb{e}_i )=-\log\left( \esp{e^{-\lambda Y_i(t)}}\right)$ in Definition \ref{lapexpdef}; we have that
\begin{align*}
&\esp{Y_i(t)}= \restr{ -\frac{
\partial}{\partial \lambda}
\esp{e^{-\lambda Y_i(t)} }
}{\lambda=0}
= \restr{ -\frac{
\partial}{\partial \lambda}
e^{-\psi_t( \lambda \pmb{e}_i ) }
}{\lambda=0}
= \restr{ -\frac{
\partial}{\partial \lambda}
e^{-\esp{ \, \psi_t^\star( \lambda W_i ) } }
}{\lambda=0}
\\ & = \restr{ 
e^{-\esp{ \, \psi_t^\star( \lambda W_i ) } }
\esp{ \, (\psi_t^\star)'( \lambda W_i )W_i }
}{\lambda=0}= (\psi_t^\star)'(0) \esp{W_i}
\end{align*}
For $0<p<1$ we can use Theorem 1 in \cite{wolfe} to obtain that
\begin{align*}
&\esp{\left(Y_i(t)\right)^p}=
\frac{p}{\Gamma(1-p)}
\int_0^\infty 
\frac{1-e^{-\psi_t(u\pmb{e}_i)}}{u^{p+1}} \d u
=\frac{p}{\Gamma(1-p)}
\int_0^\infty 
\frac{1-e^{-\esp{\psi^\star_t(u W_i)}}}{u^{p+1}} \d u
\end{align*}
We observe that 
\begin{align*}
&\esp{Y_i^2(t)} = 
\restr{
\frac{\partial^2}{\partial \lambda \partial \lambda }
\esp{e^{-\lambda Y_i(t)}}
}{
\lambda  = 0
}
=\restr{
\frac{\partial^2}{\partial \lambda \partial \lambda }
e^{-\psi_t (\lambda \pmb{e}_i) }
}{
\lambda  = 0
} =
\restr{
\frac{\partial^2}{\partial \lambda \partial \lambda }
e^{-\esp{ \psi_t^\star (\lambda W_i) } }
}{
\lambda  = 0
}
\\&=
\restr{
\frac{\partial}{\partial \lambda }
e^{-\esp{ \psi_t^\star (\lambda W_i) } }
(-1)\esp{ (\psi_t^\star)' (\lambda W_i)W_i }
}{
\lambda  = 0
}
\\&=
\restr{
\left(
e^{-\esp{ \psi_t^\star (\lambda W_i) } }
(\esp{ (\psi_t^\star)' (\lambda W_i)W_i })^2
-e^{-\esp{ \psi_t^\star (\lambda W_i) } }
\esp{ (\psi_t^\star)'' (\lambda W_i)W_i^2 }
\right)
}{
\lambda  = 0
}
\\&=
\left( (\psi_t^\star)'(0) \right)^2
\esp{ W_i }^2
- (\psi_t^\star)'' (0)\esp{ W_i^2 }.
\end{align*}
It follows that
\begin{align*}
\var{Y_i(t)}
&=\esp{Y_i^2(t)}- \esp{Y_i(t)}^2
= \left( (\psi_t^\star)'(0) \right)^2
\esp{ W_i }^2
- (\psi_t^\star)'' (0)\esp{ W_i^2 }
- \left( (\psi_t^\star)'(0) \right)^2 \esp{W_i}^2
\\&= - (\psi_t^\star)'' (0)\esp{ W_i^2 }
\end{align*}
For $i,j\in \left\{ 1, \ldots , n\right\}$, $i\neq j$ observe that
\begin{align*}
&\esp{Y_i(t)Y_j(t)} = 
\restr{
\frac{\partial^2}{\partial \lambda_j \partial \lambda_i}
\esp{e^{-\lambda_i Y_i(t)-\lambda_j Y_j(t)}}
}{
\lambda_i = \lambda_j = 0
}
=
\restr{
\frac{\partial^2}{\partial \lambda_j \partial \lambda_i}
e^{-\psi_t (\lambda_i \pmb{e}_i + \lambda_j \pmb{e}_j )}
}{
\lambda_i = \lambda_j = 0
}
\\&=
\restr{
\frac{\partial^2}{\partial \lambda_j \partial \lambda_i}
e^{-\esp{ \psi_t^\star (\lambda_i W_i + \lambda_j W_j )}}
}{
\lambda_i = \lambda_j = 0
}
=
\restr{
\frac{\partial}{\partial \lambda_j }
e^{-\esp{ \psi_t^\star (\lambda_i W_i + \lambda_j W_j )}}
(-1)\esp{ (\psi_t^\star)' (\lambda_i W_i + \lambda_j W_j )W_i}
}{
\lambda_i = \lambda_j = 0
}
\\&=
\left(
e^{-\esp{ \psi_t^\star (\lambda_i W_i + \lambda_j W_j )}}
\esp{ (\psi_t^\star)' (\lambda_i W_i + \lambda_j W_j )W_j}
\esp{ (\psi_t^\star)' (\lambda_i W_i + \lambda_j W_j )W_i}
\right.
\\&\hphantom{=}\restr{ \left. - e^{-\esp{ \psi_t^\star (\lambda_i W_i + \lambda_j W_j )}}\esp{ (\psi_t^\star)'' (\lambda_i W_i + \lambda_j W_j )W_iW_j}
\right)
}{
\lambda_i = \lambda_j = 0
} 
\\&=
\left( (\psi_t^\star)' (0) \right)^2
\esp{  W_j} \esp{ W_i}
-\esp{ (\psi_t^\star)'' (0) W_iW_j}
\end{align*}
We get that
\begin{align*}
\cov{Y_i(t),Y_j(t)}&= \esp{Y_i(t)Y_j(t)}  - \esp{Y_i(t)}\esp{Y_j(t)}
\\&=   \left( (\psi_t^\star)' (0) \right)^2
\esp{  W_j} \esp{ W_i}
-\esp{ (\psi_t^\star)'' (0) W_iW_j} 
- \left( (\psi_t^\star)'(0)  \right)^2 \esp{W_i}
\esp{W_j}
\\&= -(\psi_t^\star)'' (0)  \esp{ W_iW_j} 
\end{align*}
It follows that
\begin{align*}
\corr{Y_i(t),Y_j(t)}&=
\frac{
\cov{Y_i(t),Y_j(t)}
}{\sqrt{\var{Y_i(t)}} \sqrt{\var{Y_j(t)}}}=
\frac{
-(\psi_t^\star)'' (0)\esp{ W_iW_j} 
}{
\sqrt{ \left( (\psi_t^\star)'' (0) \right)^2 \esp{ W_i^2 }
\esp{ W_j^2 }
}
}
\\&=
\frac{
\esp{ W_iW_j} 
}{
\sqrt{ \esp{ W_i^2 }\esp{ W_j^2 }
}
}
\end{align*}

\noindent\subsection*{Proof of Theorem \ref{claylevycopteo} }\label{proofteo1}
Proof of a) 
\\
The proof strategy is to use \eqref{sklarlevycop} in order to obtain the corresponding L\'evy copula. As first step, we obtain the bivariate tail integral associated to $\rho_{\sigma,K,\pmb{\alpha},\pmb{\beta}}$ as in the hypothesis; we denote this tail integral by $U$.
\begin{align*}
U(y_1,y_2)&=\int_{y_1}^\infty \int_{y_2}^\infty\frac{\sigma K \beta_1^{\alpha_1}\beta_2^{\alpha_2} 
\Gamma(\alpha_1+\alpha_2+\sigma)
s_1^{\alpha_1 - 1} s_2^{\alpha_2-1}
(\beta_1 s_1+\beta_2 s_2)^{-\alpha_1-\alpha_2-\sigma}}{
\Gamma(\alpha_1)\Gamma(\alpha_2)} 
\mathrm{d}s_1\mathrm{d}s_2
\\
&=\int_{\beta_1 y_1}^\infty \int_{\beta_2 y_2}^\infty\frac{\sigma K 
\Gamma(\alpha_1+\alpha_2+\sigma)
s_1^{\alpha_1 - 1} s_2^{\alpha_2-1}
(s_1+s_2)^{-\alpha_1-\alpha_2-\sigma}}{
\Gamma(\alpha_1)\Gamma(\alpha_2)} 
\mathrm{d}s_1\mathrm{d}s_2
\end{align*}
We consider the change of variable
\begin{align*}
\pmb{h}(s_1,s_2) &= (s_1+s_2, s_1/(s_1+s_2) ) = (\rho,z_1) 
\\
\mathrm{d}\rho \mathrm{d}z_1 &= \left|\det ( \frac{\mathrm{d}\pmb{h}}{\mathrm{d}\pmb{s}} ) \right|\mathrm{d}s_1 \mathrm{d}s_2 = (s_1+s_2)^{-1} \mathrm{d}s_1 \mathrm{d}s_2 
\end{align*}
so
\begin{align*}
&U(y_1,y_2)=\frac{\sigma K \Gamma(\alpha_1+\alpha_2+\sigma)}{\Gamma(\alpha_1)\Gamma(\alpha_2)}
\\
&\times 
\int_{h\left( \left\{s_1,s_2 \, : \, \beta_1 y_1\leq s_1 \, , \, \beta_2 y_2\leq s_2 \right\} \right)}
z_1^{\alpha_1 -1}(1-z_1)^{\alpha_2 -1}\rho^{-\sigma -1}\mathrm{d}\rho \mathrm{d}z_1.
\end{align*}
Throughout the proof, we denote with $c_{\sigma,\pmb{\alpha}}$ the following quantity
$$
c_{\sigma, \pmb{\alpha}} = \frac{K \Gamma(\alpha_1+\alpha_2+\sigma)}{\Gamma(\alpha_1)\Gamma(\alpha_2)}
$$
For the integration region, we consider the curves
\begin{align*}
\hat{\omega}(\hat{t}) & =h(\beta_1 y_1,\beta_2 y_2+\hat{t})=(\beta_1 y_1+\beta_2 y_2+\hat{t}, \beta_1 y_1/(\beta_1 y_1+\beta_2 y_2+\hat{t}))
\\
\hat{\gamma}(\hat{t})&=h(\beta_1 y_1+\hat{t},y_2)=(\beta_1 y_1 +\beta_2 y_2+\hat{t},(\beta_1 y_1+\hat{t})/(\beta_1 y_1+\beta_2 y_2+\hat{t}))
\end{align*}
with $\hat{t}\geq 0$;
so for $t_1 = \beta_1 y_1/(\beta_1 y_1 + \beta_2 y_2+\hat{t})$ and $t_2 = (\beta_1 y_1+\hat{t})/(\beta_1 y_1 + \beta_2 y_2+\hat{t})$ we can get the reparametrized curves $
\omega(t_1)= (\beta_1 y_1/t_1,t_1)$ and $\gamma(t_2)= (\beta_2 y_2/(1-t_2),t_2)$ to delimit the integration area, hence using Fubini theorem
\begin{align*}
U(y_1,y_2)&=
c_{\sigma, \pmb{\alpha}}\left(
\int_0^{\frac{\beta_1 y_1}{\beta_1 y_1 +\beta_2 y_2}} \int_{\beta_1 y_1/z_1}^\infty
z_1^{\alpha_1 -1}(1-z_1)^{\alpha_2 -1}\sigma\rho^{-\sigma -1}
\d \rho \d z_1  \right.
\\
&\hphantom{=} \left. +
\int_{\frac{\beta_1 y_1}{\beta_1 y_1 +\beta_2 y_2}}^1 \int_{\beta_2 y_2/(1-z_1)}^\infty
z_1^{\alpha_1 -1}(1-z_1)^{\alpha_2 -1}\sigma\rho^{-\sigma -1}
\d \rho \d z_1 \right)
\\
&=
c_{\sigma, \pmb{\alpha}}\left(
\int_0^{\frac{\beta_1 y_1}{\beta_1 y_1 +\beta_2 y_2}}
z_1^{\alpha_1 + \sigma -1}(1-z_1)^{\alpha_2 -1}
(\beta_1 y_1)^{-\sigma} 
\d z_1  \right.
\\
&\hphantom{=} \left. + 
\int_{\frac{\beta_1 y_1}{\beta_1 y_1 +\beta_2 y_2}}^1 
z_1^{\alpha_1 -1}(1-z_1)^{\alpha_2 + \sigma -1} 
(\beta_2 y_2 )^{-\sigma}
\d z_1 \right)
\\
&=
c_{\sigma, \pmb{\alpha}}\left(
\int_0^{\frac{\beta_1 y_1}{\beta_1 y_1 +\beta_2 y_2}}
z_1^{\alpha_1 + \sigma -1}(1-z_1)^{\alpha_2 -1}
(\beta_1 y_1)^{-\sigma} 
\d z_1  \right.
\\
&\hphantom{=} \left. + 
\int_0^{\frac{\beta_2 y_2}{\beta_1 y_1 +\beta_2 y_2}} 
z_1^{\alpha_2+\sigma -1}(1-z_1)^{\alpha_1 -1} 
(\beta_2 y_2 )^{-\sigma}
\d z_1 \right)
\end{align*}
The above expression can be evaluated in terms of cumulative distribution functions of a Beta$(\alpha,\beta)$ random variable which we write as the regularized incomplete beta function $I(x,\alpha,\beta)$. Let $B$ be the beta function $B(a,b)=\Gamma(a)\Gamma(b)/\Gamma(a+b)$, thus
\begin{align*}
&U(y_1,y_2)=c_{\sigma, \pmb{\alpha}}\left(
(\beta_1 y_1)^{-\sigma}B(\alpha_1+\sigma,\alpha_2)I\left( \frac{\beta_1 y_1}{\beta_1 y_1 +\beta_2 y_2},\alpha_1+\sigma,\alpha_2\right) \right.
\\
& \left. \hphantom{U(y_1,y_2)=} + 
(\beta_2 y_2)^{-\sigma}B(\alpha_1,\alpha_2+\sigma)I\left( \frac{\beta_2 y_2}{\beta_1 y_1 +\beta_2 y_2},\alpha_2+\sigma,\alpha_1\right)
\right)
\end{align*}
\begin{align*}
\hphantom{U(y_1,y_2)}&=
\frac{K\Gamma(\alpha_1+\sigma)
(\beta_1 y_1)^{-\sigma}}{\Gamma(\alpha_1)}I\left( \frac{\beta_1 y_1}{\beta_1 y_1 +\beta_2 y_2},\alpha_1+\sigma,\alpha_2\right)
\\
\hphantom{U(y_1,y_2)=} & + 
\frac{ K \Gamma(\alpha_2 +\sigma)
(\beta_2 y_2)^{-\sigma}}{\Gamma(\alpha_2)}
I\left( \frac{\beta_2 y_2}{\beta_1 y_1 +\beta_2 y_2},\alpha_2+\sigma,\alpha_1\right)
\end{align*}
To get the copula we evaluate the above tail integral in 
$$\left(U_1^{-1}(y_1),U^{-1}(y_2)\right)=
\left(
\left( \frac{\Gamma(\alpha_1)y_1}{K\beta_1^{-\sigma}\Gamma(\alpha_1+\sigma)} \right)^{-\frac{1}{\sigma}}
,
\left( \frac{\Gamma(\alpha_2)y_2}{K\beta_2^{-\sigma}\Gamma(\alpha_2+\sigma)} \right)^{-\frac{1}{\sigma}}
\right).
$$
So
\begin{align*}
\mathcal{C}(s_1, s_2) &=
s_1\,
I\left(
\frac{ \left( \frac{\Gamma(\alpha_1+\sigma)}{\Gamma(\alpha_1)s_1} \right)^{\frac{1}{\sigma}} }{ 
\left( \frac{\Gamma(\alpha_1+\sigma)}{\Gamma(\alpha_1)s_1} \right)^{\frac{1}{\sigma}} +
\left( \frac{\Gamma(\alpha_2+\sigma)}{\Gamma(\alpha_2)s_2} \right)^{\frac{1}{\sigma}}
},\alpha_1+\sigma,\alpha_2 \right)
\\
&\hphantom{=}+
s_2\,
I\left(
\frac{ \left( \frac{\Gamma(\alpha_2+\sigma)}{\Gamma(\alpha_2)s_2} \right)^{\frac{1}{\sigma}} }{ 
\left( \frac{\Gamma(\alpha_1+\sigma)}{\Gamma(\alpha_1)s_1} \right)^{\frac{1}{\sigma}} +
\left( \frac{\Gamma(\alpha_2+\sigma)}{\Gamma(\alpha_2)s_2} \right)^{\frac{1}{\sigma}}
},\alpha_2 +\sigma, \alpha_1 \right)
\end{align*}
\\
Proof of b)
\\
We perform a constructive proof by using Theorem 6.3 in \cite{tankbook} to show that for $\sigma >1$ a vector of subordinators which has $\mathcal{C}_{\sigma,\pmb{\alpha}}$ as its associated L\'evy copula can be given. So $\mathcal{C}_{\sigma,\pmb{\alpha}}$ will be a L\'evy copula also for $\sigma >1$.
Using Theorem 6.3 in \cite{tankbook}, we need to show that $F_{S_1|S_2=s_2}(s_1) = 
\frac{\partial}{\partial s_2} \mathcal{C}_{\sigma,\pmb{\alpha}}(s_1, s_2)$ and $F_{S_2|S_1=s_1}(s_2) = \frac{\partial}{\partial s_1} \mathcal{C}_{\sigma,\pmb{\alpha}}(s_1, s_2)$ are cumulative distribution functions. We observe that
\begin{align*}
& \frac{\partial}{\partial s_1} \mathcal{C}_{\sigma , \pmb{\alpha}}(s_1, s_2)  = 
\frac{\Gamma(\alpha_1+\alpha_2+\sigma)}{\Gamma(\alpha_1+\sigma)\Gamma(\alpha_2)}
\int_0^{
\frac{ \left( \frac{\Gamma(\alpha_1+\sigma)}{\Gamma(\alpha_1)s_1} \right)^{\frac{1}{\sigma}} }{ 
\left( \frac{\Gamma(\alpha_1+\sigma)}{\Gamma(\alpha_1)s_1} \right)^{\frac{1}{\sigma}} +
\left( \frac{\Gamma(\alpha_2+\sigma)}{\Gamma(\alpha_2)s_2} \right)^{\frac{1}{\sigma}} }
}z^{\alpha_1+\sigma -1}(1-z)^{\alpha_2-1}\d z
\\
&-
\frac{\Gamma(\alpha_1+\alpha_2+\sigma)}{\sigma\Gamma(\alpha_1+\sigma)\Gamma(\alpha_2)}
\frac{
\left( \frac{\Gamma(\alpha_1 +\sigma)}{\Gamma(\alpha_1)s_1}  \right)^{\frac{\alpha_1+\sigma}{\sigma}}
\left( \frac{\Gamma(\alpha_2 +\sigma)}{\Gamma(\alpha_2)s_2}  \right)^{\frac{\alpha_2}{\sigma}}
}{
\left(
\left( \frac{\Gamma(\alpha_1 +\sigma)}{\Gamma(\alpha_1)s_1}  \right)^{\frac{1}{\sigma}} +
\left( \frac{\Gamma(\alpha_2 +\sigma)}{\Gamma(\alpha_2)s_2}  \right)^{\frac{1}{\sigma}}
\right)^{\alpha_1 +\alpha_2 +\sigma}
}
\\
&+
\frac{\Gamma(\alpha_1+\alpha_2+\sigma)s_2}{\sigma\Gamma(\alpha_2+\sigma)\Gamma(\alpha_1)s_1}
\frac{
\left( \frac{\Gamma(\alpha_1 +\sigma)}{\Gamma(\alpha_1)s_1}  \right)^{\frac{\alpha_1}{\sigma}}
\left( \frac{\Gamma(\alpha_2 +\sigma)}{\Gamma(\alpha_2)s_2}  \right)^{\frac{\alpha_2+\sigma}{\sigma}}
}{
\left(
\left( \frac{\Gamma(\alpha_1 +\sigma)}{\Gamma(\alpha_1)s_1}  \right)^{\frac{1}{\sigma}} +
\left( \frac{\Gamma(\alpha_2 +\sigma)}{\Gamma(\alpha_2)s_2}  \right)^{\frac{1}{\sigma}}
\right)^{\alpha_1 +\alpha_2 +\sigma}
}
\\
&=
\frac{\Gamma(\alpha_1+\alpha_2+\sigma)}{
\Gamma(\alpha_1+\sigma)\Gamma(\alpha_2)}
\int_0^{
\frac{ \left( \frac{\Gamma(\alpha_1+\sigma)}{\Gamma(\alpha_1)s_1} \right)^{\frac{1}{\sigma}} }{ 
\left( \frac{\Gamma(\alpha_1+\sigma)}{\Gamma(\alpha_1)s_1} \right)^{\frac{1}{\sigma}} +
\left( \frac{\Gamma(\alpha_2+\sigma)}{\Gamma(\alpha_2)s_2} \right)^{\frac{1}{\sigma}} }
}z^{\alpha_1+\sigma -1}(1-z)^{\alpha_2-1}\d z
\\
&-
\frac{\Gamma(\alpha_1+\alpha_2+\sigma)}{\sigma\Gamma(\alpha_1)\Gamma(\alpha_2)s_1}
\frac{
\left( \frac{\Gamma(\alpha_1 +\sigma)}{\Gamma(\alpha_1)s_1}  \right)^{\frac{\alpha_1}{\sigma}}
\left( \frac{\Gamma(\alpha_2 +\sigma)}{\Gamma(\alpha_2)s_2}  \right)^{\frac{\alpha_2}{\sigma}}
}{
\left(
\left( \frac{\Gamma(\alpha_1 +\sigma)}{\Gamma(\alpha_1)s_1}  \right)^{\frac{1}{\sigma}} +
\left( \frac{\Gamma(\alpha_2 +\sigma)}{\Gamma(\alpha_2)s_2}  \right)^{\frac{1}{\sigma}}
\right)^{\alpha_1 +\alpha_2 +\sigma}
}
\\
&+
\frac{\Gamma(\alpha_1+\alpha_2+\sigma)}{\sigma\Gamma(\alpha_1)\Gamma(\alpha_2)s_1}
\frac{
\left( \frac{\Gamma(\alpha_1 +\sigma)}{\Gamma(\alpha_1)s_1}  \right)^{\frac{\alpha_1}{\sigma}}
\left( \frac{\Gamma(\alpha_2 +\sigma)}{\Gamma(\alpha_2)s_2}  \right)^{\frac{\alpha_2}{\sigma}}
}{
\left(
\left( \frac{\Gamma(\alpha_1 +\sigma)}{\Gamma(\alpha_1)s_1}  \right)^{\frac{1}{\sigma}} +
\left( \frac{\Gamma(\alpha_2 +\sigma)}{\Gamma(\alpha_2)s_2}  \right)^{\frac{1}{\sigma}}
\right)^{\alpha_1 +\alpha_2 +\sigma}
}
\\
&=
\frac{\Gamma(\alpha_1+\alpha_2+\sigma)}{
\Gamma(\alpha_1+\sigma)\Gamma(\alpha_2)}
\int_0^{
\frac{ \left( \frac{\Gamma(\alpha_1+\sigma)}{\Gamma(\alpha_1)s_1} \right)^{\frac{1}{\sigma}} }{ 
\left( \frac{\Gamma(\alpha_1+\sigma)}{\Gamma(\alpha_1)s_1} \right)^{\frac{1}{\sigma}} +
\left( \frac{\Gamma(\alpha_2+\sigma)}{\Gamma(\alpha_2)s_2} \right)^{\frac{1}{\sigma}} }
}z^{\alpha_1+\sigma -1}(1-z)^{\alpha_2-1}\d z
\end{align*}
and similarly
\begin{align*}
\frac{\partial}{\partial s_2} \mathcal{C}_{\sigma , \pmb{\alpha}}(s_1, s_2)  = 
\frac{\Gamma(\alpha_1+\alpha_2+\sigma)}{
\Gamma(\alpha_2+\sigma)\Gamma(\alpha_1)}
\int_0^{
\frac{ \left( \frac{\Gamma(\alpha_2+\sigma)}{\Gamma(\alpha_2)s_2} \right)^{\frac{1}{\sigma}} }{ 
\left( \frac{\Gamma(\alpha_1+\sigma)}{\Gamma(\alpha_1)s_1} \right)^{\frac{1}{\sigma}} +
\left( \frac{\Gamma(\alpha_2+\sigma)}{\Gamma(\alpha_2)s_2} \right)^{\frac{1}{\sigma}} }
}z^{\alpha_2+\sigma -1}(1-z)^{\alpha_1-1}\d z
\end{align*}
Either by differentiating the above expressions or by using Theorem \ref{sklarlevycop} we can obtain that
\begin{align*}
& \frac{\partial^2}{\partial s_2 \partial s_1} \mathcal{C}_{\sigma , \pmb{\alpha}}(s_1, s_2) =
\frac{\Gamma(\alpha_1+\alpha_2+\sigma)}{\sigma
\Gamma(\alpha_1)\Gamma(\alpha_2)s_1 s_2}
\frac{
\left( \frac{\Gamma(\alpha_1 +\sigma)}{\Gamma(\alpha_1)s_1}  \right)^{\frac{\alpha_1}{\sigma}}
\left( \frac{\Gamma(\alpha_2 +\sigma)}{\Gamma(\alpha_2)s_2}  \right)^{\frac{\alpha_2}{\sigma}}
}{
\left(
\left( \frac{\Gamma(\alpha_1 +\sigma)}{\Gamma(\alpha_1)s_1}  \right)^{\frac{1}{\sigma}} +
\left( \frac{\Gamma(\alpha_2 +\sigma)}{\Gamma(\alpha_2)s_2}  \right)^{\frac{1}{\sigma}}
\right)^{\alpha_1 +\alpha_2 +\sigma}
}
\end{align*}
So for any $\sigma > 0$ $F_{S_1|S_2=s_2}(s_1)$ and $F_{S_2|S_1=s_1}(s_2)$ are monotone functions. It suffices to check that
\begin{align*}
\lim_{s_2 \to 0 } F_{S_2|S_1=s_1}(s_2) & =
\lim_{s_2 \to 0 }
\frac{\partial}{\partial s_1} \mathcal{C}_{\sigma , \pmb{\alpha}}(s_1, s_2) = 0
\end{align*}
and
\begin{align*}
\lim_{s_2 \to \infty } F_{S_2|S_1=s_1}(s_2) & = \lim_{s_2 \to \infty } 
\frac{\partial}{\partial s_1} \mathcal{C}_{\sigma , \pmb{\alpha}}(s_1, s_2) = 1,
\end{align*}
having the case for $F_{S_1|S_2=s_2}(s_1)$ being analogous.
For the first limit we use the monotonous convergence theorem to obtain that 
\begin{align*}
\lim_{s_2 \to 0 } \frac{\partial}{\partial s_1} \mathcal{C}_{\sigma , \pmb{\alpha}}(s_1, s_2)
&=
\lim_{s_2 \to 0 } 
\frac{\Gamma(\alpha_1+\alpha_2+\sigma)}{
\Gamma(\alpha_1+\sigma)\Gamma(\alpha_2)}
\int_0^{
\frac{ \left( \frac{\Gamma(\alpha_1+\sigma)}{\Gamma(\alpha_1)s_1} \right)^{\frac{1}{\sigma}} }{ 
\left( \frac{\Gamma(\alpha_1+\sigma)}{\Gamma(\alpha_1)s_1} \right)^{\frac{1}{\sigma}} +
\left( \frac{\Gamma(\alpha_2+\sigma)}{\Gamma(\alpha_2)s_2} \right)^{\frac{1}{\sigma}} }
}z^{\alpha_1+\sigma -1}(1-z)^{\alpha_2-1}\d z
\\
&=
\frac{\Gamma(\alpha_1+\alpha_2+\sigma)}{
\Gamma(\alpha_1+\sigma)\Gamma(\alpha_2)}
\int_0^0 z^{\alpha_1+\sigma -1}(1-z)^{\alpha_2-1}\d z
= 0
\end{align*}
So
\begin{align*}
\lim_{s_2 \to 0 } F_{S_2|S_1=s_1}(s_2)
= 0.
\end{align*}
Using the monotonous convergence theorem again we also obtain  
\begin{align*}
\lim_{s_2 \to \infty } \frac{\partial}{\partial s_1} \mathcal{C}_{\sigma , \pmb{\alpha}}(s_1, s_2)
&=
\lim_{s_2 \to \infty } 
\frac{\Gamma(\alpha_1+\alpha_2+\sigma)}{
\Gamma(\alpha_1+\sigma)\Gamma(\alpha_2)}
\int_0^{
\frac{ \left( \frac{\Gamma(\alpha_1+\sigma)}{\Gamma(\alpha_1)s_1} \right)^{\frac{1}{\sigma}} }{ 
\left( \frac{\Gamma(\alpha_1+\sigma)}{\Gamma(\alpha_1)s_1} \right)^{\frac{1}{\sigma}} +
\left( \frac{\Gamma(\alpha_2+\sigma)}{\Gamma(\alpha_2)s_2} \right)^{\frac{1}{\sigma}} }
}z^{\alpha_1+\sigma -1}(1-z)^{\alpha_2-1}\d z
\\
&=
\frac{\Gamma(\alpha_1+\alpha_2+\sigma)}{
\Gamma(\alpha_1+\sigma)\Gamma(\alpha_2)}
\int_0^1 z^{\alpha_1+\sigma -1}(1-z)^{\alpha_2-1}\d z
= 1
\end{align*}
So
\begin{align*}
\lim_{s_2 \to \infty } F_{S_2|S_1=s_1}(s_2) = 1
\end{align*}
As these limits do not depend on what values $\sigma$ takes in $(0,\infty )$ we conclude by using Theorem 6.3 in \cite{tankbook} that we can construct a subordinator with the desired L\'evy copula for any $\sigma \in (0, \infty )$.

\subsection*{Proof of Theorem \ref{limbehavteo} }\label{prooflimbehav}

We start with the proof for $\sigma\to 0$, i.e. $\theta = \frac{1}{\sigma} \to \infty$. Observe that 
$$
\lim_{\sigma\to 0}\frac{a^{\frac{1}{\sigma}}}{a^{\frac{1}{\sigma}}+b^{\frac{1}{\sigma}}}
=
\lim_{\sigma\to 0}\frac{1}{1+\left( \frac{b}{a} \right)^{\frac{1}{\sigma}}}
$$
so when $b<a$ the limit is $1$ and when $a<b$ the limit is $0$. On the other hand $\lim_{\sigma \to 0} \frac{\Gamma(a+\sigma)}{\Gamma(a)}=1$ for any $a>0$ so there exists $\delta>0$ such that for $0<\sigma<\delta$ we have that i) $s_1<s_2 \implies \frac{\Gamma(\alpha_2+\sigma)}{\Gamma(\alpha_2)s_2}<\frac{\Gamma(\alpha_1+\sigma)}{\Gamma(\alpha_1)s_1}$ and ii) $s_2<s_1 \implies \frac{\Gamma(\alpha_1+\sigma)}{\Gamma(\alpha_1)s_1}<\frac{\Gamma(\alpha_2+\sigma)}{\Gamma(\alpha_2)s_2}$.
Without loss of generality we assume that $s_1<s_2$, the case $s_2<s_1$ being treated analogously. We consider $0<\sigma <\delta$ and use the bounded convergence theorem to see that
\begin{align*}
& \lim_{\sigma \to 0 } \mathcal{C}_{\sigma,\pmb{\alpha}}(s_1,s_2)
= 
\lim_{\sigma \to 0 }
\left( 
s_1
\frac{\Gamma(\alpha_1+\alpha_2+\sigma)}{
\Gamma(\alpha_1+\sigma)\Gamma(\alpha_2)}
\int_0^{
\frac{ \left( \frac{\Gamma(\alpha_1+\sigma)}{\Gamma(\alpha_1)s_1} \right)^{\frac{1}{\sigma}} }{ 
\left( \frac{\Gamma(\alpha_1+\sigma)}{\Gamma(\alpha_1)s_1} \right)^{\frac{1}{\sigma}} +
\left( \frac{\Gamma(\alpha_2+\sigma)}{\Gamma(\alpha_2)s_2} \right)^{\frac{1}{\sigma}} }
}z^{\alpha_1+\sigma -1}(1-z)^{\alpha_2-1}\d z
\right.
\\ & \left. 
+ s_2
\frac{\Gamma(\alpha_1+\alpha_2+\sigma)}{
\Gamma(\alpha_2+\sigma)\Gamma(\alpha_1)}
\int_0^{
\frac{ \left( \frac{\Gamma(\alpha_2+\sigma)}{\Gamma(\alpha_2)s_2} \right)^{\frac{1}{\sigma}} }{ 
\left( \frac{\Gamma(\alpha_1+\sigma)}{\Gamma(\alpha_1)s_1} \right)^{\frac{1}{\sigma}} +
\left( \frac{\Gamma(\alpha_2+\sigma)}{\Gamma(\alpha_2)s_2} \right)^{\frac{1}{\sigma}} }
}z^{\alpha_2+\sigma -1}(1-z)^{\alpha_1-1}\d z
\right)
\\
=
& s_1
\frac{\Gamma(\alpha_1+\alpha_2+\sigma)}{
\Gamma(\alpha_1+\sigma)\Gamma(\alpha_2)}
\int_0^1 z^{\alpha_1+\sigma -1}(1-z)^{\alpha_2-1}\d z 
+ s_2
\frac{\Gamma(\alpha_1+\alpha_2+\sigma)}{
\Gamma(\alpha_2+\sigma)\Gamma(\alpha_1)}
\int_0^0 z^{\alpha_2+\sigma -1}(1-z)^{\alpha_1-1}\d z
= s_1.
\end{align*}
Similarly if $s_2<s_1$, $\lim_{\sigma \to 0 } \mathcal{C}_{\sigma,\pmb{\alpha}}(s_1,s_2)=s_2$ so by continuity $\lim_{\sigma \to 0 } \mathcal{C}_{\sigma,\pmb{\alpha}}(s_1,s_2)=\min\left\{ s_1 , s_2 \right\}$. 
\\
We continue the proof for $\sigma\to \infty$, i.e. $\theta = \frac{1}{\sigma} \to 0$. Observe that $\mathcal{C}_{\sigma,\pmb{\alpha}}(s_1,s_2)=0$ if either $s_1=0$ or $s_2 = 0$. On the other hand, by continuity
$$
\lim_{\sigma\to \infty }\frac{a^{\frac{1}{\sigma}}}{a^{\frac{1}{\sigma}}+b^{\frac{1}{\sigma}}} = \frac{1}{2}
$$
for $a,b>0$. So there exists $n_0 >0$ such that for $n>n_0$ and $s_1,s_2 > 0$ we have that
\begin{align*}
&0\leq \lim_{\sigma \to \infty } \mathcal{C}_{\sigma,\pmb{\alpha}}(s_1,s_2)
= 
\lim_{\sigma \to \infty }
\left( 
s_1
\frac{\Gamma(\alpha_1+\alpha_2+\sigma)}{
\Gamma(\alpha_1+\sigma)\Gamma(\alpha_2)}
\int_0^{
\frac{ \left( \frac{\Gamma(\alpha_1+\sigma)}{\Gamma(\alpha_1)s_1} \right)^{\frac{1}{\sigma}} }{ 
\left( \frac{\Gamma(\alpha_1+\sigma)}{\Gamma(\alpha_1)s_1} \right)^{\frac{1}{\sigma}} +
\left( \frac{\Gamma(\alpha_2+\sigma)}{\Gamma(\alpha_2)s_2} \right)^{\frac{1}{\sigma}} }
}z^{\alpha_1+\sigma -1}(1-z)^{\alpha_2-1}\d z
\right.
\\ & \left. 
+ s_2
\frac{\Gamma(\alpha_1+\alpha_2+\sigma)}{
\Gamma(\alpha_2+\sigma)\Gamma(\alpha_1)}
\int_0^{
\frac{ \left( \frac{\Gamma(\alpha_2+\sigma)}{\Gamma(\alpha_2)s_2} \right)^{\frac{1}{\sigma}} }{ 
\left( \frac{\Gamma(\alpha_1+\sigma)}{\Gamma(\alpha_1)s_1} \right)^{\frac{1}{\sigma}} +
\left( \frac{\Gamma(\alpha_2+\sigma)}{\Gamma(\alpha_2)s_2} \right)^{\frac{1}{\sigma}} }
}z^{\alpha_2+\sigma -1}(1-z)^{\alpha_1-1}\d z
\right)
\\
<
& \lim_{\sigma \to \infty } \left( s_1
\frac{\Gamma(\alpha_1+\alpha_2+\sigma)}{
\Gamma(\alpha_1+\sigma)\Gamma(\alpha_2)}
\int_0^{\frac{1}{2}} z^{\alpha_1+\sigma -1}(1-z)^{\alpha_2-1}\d z 
+ s_2
\frac{\Gamma(\alpha_1+\alpha_2+\sigma)}{
\Gamma(\alpha_2+\sigma)\Gamma(\alpha_1)}
\int_0^{\frac{1}{2}} z^{\alpha_2+\sigma -1}(1-z)^{\alpha_1-1}\d z \right)
= 0.
\end{align*}
So $ \mathcal{C}_{\sigma,\pmb{\alpha}}(s_1,s_2)$ can only be non zero when $s_1 \to \infty $ or $s_2 \to \infty$. We say that for real functions $f,g$ $f \approx g$ as $x\to c$ if $\lim_{x\to x}\frac{f(x)}{g(x)}=1$ and observe that for $x,y,a,b,\sigma>0$, as $x\to \infty$
\begin{align*}
&x
\frac{\Gamma(a+b+\sigma)}{
\Gamma(a+\sigma)\Gamma(b)}
\int_0^{
\frac{ \left( \frac{\Gamma(a+\sigma)}{\Gamma(a)x} \right)^{\frac{1}{\sigma}} }{ 
\left( \frac{\Gamma(a+\sigma)}{\Gamma(a)x} \right)^{\frac{1}{\sigma}} +
\left( \frac{\Gamma(b+\sigma)}{\Gamma(b) y} \right)^{\frac{1}{\sigma}} }
}z^{a+\sigma -1}(1-z)^{b-1}\d z
\\
&
\approx
x 
\frac{\Gamma(a+b+\sigma)}{
\Gamma(a+\sigma)\Gamma(b)}
\left(
\frac{ \left( \frac{\Gamma(a+\sigma)}{\Gamma(a)x} \right)^{\frac{1}{\sigma}} }{ 
\left( \frac{\Gamma(a+\sigma)}{\Gamma(a)x} \right)^{\frac{1}{\sigma}} +
\left( \frac{\Gamma(b+\sigma)}{\Gamma(b) y} \right)^{\frac{1}{\sigma}} }
\right)^{a+\sigma-1}
\frac{ \left( \frac{\Gamma(a+\sigma)}{\Gamma(a)x} \right)^{\frac{1}{\sigma}} }{ 
\left( \frac{\Gamma(a+\sigma)}{\Gamma(a)x} \right)^{\frac{1}{\sigma}} +
\left( \frac{\Gamma(b+\sigma)}{\Gamma(b) y} \right)^{\frac{1}{\sigma}} }
\\
&
=
x 
\frac{\Gamma(a+b+\sigma)}{
\Gamma(a+\sigma)\Gamma(b)}
\left(
\frac{ \left( \frac{\Gamma(a+\sigma)}{\Gamma(a)x} \right)^{\frac{1}{\sigma}} }{ 
\left( \frac{\Gamma(a+\sigma)}{\Gamma(a)x} \right)^{\frac{1}{\sigma}} +
\left( \frac{\Gamma(b+\sigma)}{\Gamma(b) y} \right)^{\frac{1}{\sigma}} }
\right)^{a+\sigma}
\approx
x 
\frac{\Gamma(a+b+\sigma)}{
\Gamma(a+\sigma)\Gamma(b)}
\frac{ \left(
\frac{\Gamma(a+\sigma)}{\Gamma(a)x}
\right)^{\frac{a+\sigma}{\sigma}} }{ \left(
\frac{\Gamma(b+\sigma)}{\Gamma(b) y}
\right)^{\frac{a+\sigma}{\sigma}} }
\\
&= 
\frac{\Gamma(a+b+\sigma)}{
\Gamma(a+\sigma)\Gamma(b)}
\frac{ \left(
\frac{\Gamma(a+\sigma)}{\Gamma(a)}
\right)^{\frac{a+\sigma}{\sigma}} 
\left( \frac{1}{x} \right)^{\frac{a}{\sigma} }
}{ \left(
\frac{\Gamma(b+\sigma)}{\Gamma(b) y}
\right)^{\frac{a+\sigma}{\sigma}} }
\longrightarrow 0 \text{ as } x\to \infty.
\end{align*}
So
\begin{align*}
\lim_{\sigma \to \infty }\lim_{s_2 \to \infty}
\mathcal{C}_{\sigma,\pmb{\alpha}}(s_1,s_2)
= s_1
\end{align*}
and
\begin{align*}
\lim_{\sigma \to \infty }\lim_{s_1 \to \infty}
\mathcal{C}_{\sigma,\pmb{\alpha}}(s_1,s_2)
= s_2
\end{align*}
So we conclude
\begin{align*}
\lim_{\sigma \to \infty }
\mathcal{C}_{\sigma,\pmb{\alpha}}(s_1,s_2)
=s_1\indic_{ \left\{ s_2=\infty \right\} }
+ s_2 \indic_{ \left\{ s_1=\infty \right\} }.
\end{align*}
\noindent\subsection*{Proof of Theorem \ref{cormcopulateo} }
We use the Sklar theorem for L\'evy copulas, \eqref{sklarlevycop} to prove the statement.
For the tail integral we have by definition that
\begin{align*}
U(y_1, \ldots , y_d)
&=
\int_0^\infty \int_{y_1}^\infty \cdots \int_{y_d}^\infty
z^{-d}
h\left( 
\frac{ s_1}{z} , \ldots , \frac{ s_d}{z}
\right) 
\d \pmb{s}
\rho^\star ( \d z )
\\
&= 
\int_0^\infty \int_{\frac{y_1}{z}}^\infty \cdots \int_{\frac{y_d}{z}}^\infty
h\left( 
u_1 , \ldots , u_d
\right) 
\d \pmb{u}
\rho^\star ( \d z )
\\
&= 
\int_0^\infty
S\left( 
\frac{y_1}{z}, \ldots , \frac{y_d}{z}
\right) \rho^\star  (\d z)
\\
&= 
\int_0^\infty
\hat{C}\left( 
S_1 \left( \frac{y_1}{z} \right) , \ldots , S_d \left( \frac{y_d}{z} \right)
\right)
\rho^\star ( \d z ).
\end{align*}
Where in the last equation we have used the Sklar theorem for survival copulas
\begin{align*}
S(u_1, \ldots , u_d) = \hat{C}\left(
S_1(u_1), \ldots S_d(u_d)
\right).
\end{align*}
Let $i\in \{1,\ldots ,d\}$, for the $i$-th marginal tail integral observe that if we evaluate the tail integral in in $\left( y_1^{(i)}, \ldots , y_{i-1}^{(i)}, y, y_{i+1}^{(i)},\ldots, y_d^{(i)}\right)$ with $y\in\re^+$ and $y_1^{(i)}= \cdots = y_{i-1}^{(i)}= y_{i+1}^{(i)},=\ldots = y_d^{(i)} = 0$ then as $\hat{C}$ has uniform marginals we conclude that
$$
U_i (x) = \int_{0}^\infty S_i \left( 
\frac{x}{z}
\right) \rho^\star(\d z)
$$ where $S_i$ is the $i-$th marginal survival function associated to the score distribution.
From the Sklar theorem for L\'evy copulas, Theorem \ref{sklarlevycop}, we conclude the proof.

\end{document}

%% file: definitions.tex
\newcommand{\bb}[1]{\mathbb{#1}}

\newcommand{\sip}{\bb{P}}

\newcommand{\se}{\ensuremath{\bb{E}}}

\newcommand{\re}{\ensuremath{\mathbb{R}}}

\newcommand{\prob}[1]{\ensuremath{\sip\! \left[ #1 \right]}}

\newcommand{\esp}[1]{\ensuremath{\se\! \left[ #1 \right]}}

\newcommand{\var}[1]{\ensuremath{\text{Var}\! \left( #1 \right)}}
\newcommand{\cov}[1]{\ensuremath{Cov\! \left( #1 \right)}}
\newcommand{\corr}[1]{\ensuremath{Cor\! \left( #1 \right)}}

\newcommand{\indic}{\mathbbm{1}}

\renewcommand{\d}{\ensuremath{\mathrm{d}}}

\newcommand\restr[2]{{
  \left.\kern-\nulldelimiterspace 
  #1 
  \vphantom{\big|} 
  \right|_{#2} 
  }}
  
\newcommand\eval[3]{{
  \left.\kern-\nulldelimiterspace 
  #1 
  \vphantom{\big|} 
  \right|_{#2}^{#3} 
  }}
  
\newtheorem*{teo*}{Theorem}

\newtheorem*{prop*}{Proposition}

\newtheorem*{cor*}{Corollary}

\theoremstyle{definition}
\newtheorem*{midef*}{Definition}

\newtheorem{ex}{Example}
\newtheorem*{ex*}{Example}